\documentclass[12pt,preprint]{aastex}

\shorttitle{Particle Scattering off of Dispersive Waves}
\shortauthors{Schreiner, Kilian, Spanier}

\usepackage{graphicx}
\usepackage{amsmath}
\usepackage{amssymb}
\usepackage{amsthm}
\usepackage{hyperref}

\begin{document}

\title{Particle Scattering off of Right-Handed Dispersive Waves}

\author{C. Schreiner}
\affil{Centre for Space Research, North-West University, 2520 Potchefstroom, South Africa}
\affil{Lehrstuhl f\"ur Astronomie, Universit\"at W\"urzburg, 97074 W\"urzburg, Germany}
\email{cschreiner@astro.uni-wuerzburg.de}

\author{P. Kilian}
\affil{Centre for Space Research, North-West University, 2520 Potchefstroom, South Africa}

\and

\author{F. Spanier}
\affil{Centre for Space Research, North-West University, 2520 Potchefstroom, South Africa}

\begin{abstract}

Resonant scattering of fast particles off low frequency plasma waves is a major process determining transport characteristics of energetic particles in the heliosphere and contributing to their acceleration.
Usually, only Alfv{\'e}n waves are considered for this process, although dispersive waves are also present throughout the heliosphere.

We investigate resonant interaction of energetic electrons with dispersive, right-handed waves.
For the interaction of particles and a single wave a variable transformation into the rest frame of the wave can be performed.
Here, well-established analytic models derived in the framework of magnetostatic quasi-linear theory (QLT) can be used as a reference to validate simulation results.
However, this approach fails as soon as several dispersive waves are involved.
Based on analytic solutions modeling the scattering amplitude in the magnetostatic limit, we present an approach to modify these equations for the use in the plasma frame.
Thereby we aim at a description of particle scattering in the presence of several waves.

A Particle-in-Cell (PiC) code is employed to study wave-particle scattering on a micro-physically correct level and to test the modified model equations.
We investigate the interactions of electrons at different energies (from $1\,\mathrm{keV}$ to $1\,\mathrm{MeV}$) and right-handed waves with various amplitudes.
Differences between model and simulation arise in the case of high amplitudes or several waves.
Analyzing the trajectories of single particles we find no microscopic diffusion in the case of a single plasma wave, although a broadening of the particle distribution can be observed.
\end{abstract}


\section{INTRODUCTION}
\label{sec:introduction}

Charged particles in the solar wind and the interstellar medium are deflected by magnetic fields.
One of the major influences on particle transport in the heliosphere are resonant interactions between particles and plasma waves.
Quasi-linear theory (QLT), first described by \citet{jokipii_1966}, treats these interactions as resonant scattering of particles and the magnetic fields of low frequency waves, such as Alfv{\'e}n waves.
In the framework of QLT the scattering of energetic protons and non-dispersive Alfv{\'e}n waves is often in the focus of research \citep[e.g.][]{lee_1974,schlickeiser_1989}, both because of the significant simplifications in modeling and the astrophysical relevance of the process.
With a significant part of the spectrum of plasma waves in the vicinity of the Sun being Alfv{\'e}n waves, these waves are believed to be a major source of energy for the acceleration of protons.
This includes both the heating of a whole population of protons via the cyclotron and Landau resonances \citep{cranmer_2003,chandran_2010} and the further acceleration of already fast particles  based on the effect of magnetic acceleration proposed by \citet{fermi_1949}.
With regard to energetic particles, the latter case is the more interesting one, since high-energy protons in the solar wind are believed to be accelerated in shock fronts in the inner heliosphere \citep{krymskii_1977}.

Collisionless shock fronts, e.g. shocks accompanying solar flares or interplanetary traveling shocks, may accelerate particles which cross the shock front \citep[for review articles see e.g.][]{jones_1991,reames_1999}.
However, the acceleration mechanism only works on suprathermal particles which may be particles in the high energy tail of the thermal distribution \citep{gosling_1981} or otherwise pre-accelerated particles.
The energy gain of the particle increases with the number of shock crossings, until the particle is fast enough to be able to escape.
For a large enough number of energetic particles in the vicinity of the shock front self-generated waves play an important role in the further acceleration of the particles \citep{lee_1983,zank_2000,vainio_2007,ng_2008}.
Alfv\'en waves may be generated by the fast streaming particles, which then help to trap those particles near the shock, thus enabling them to cross the shock front more often before they escape.
An additional boost to acceleration efficiency can be provided if the shock normal is quasi-perpendicular to the average magnetic field and the so-called shock-drift-acceleration of energetic particles is most efficient \citep{jokipii_1987}.

Although the interaction of protons and non-dispersive Alfv{\'e}n waves might be the most important mechanism, resonant scattering of fast particles and waves of different kinds still occurs.
An obvious possibility is the scattering of protons (electrons) and waves in the dispersive regime of the Alfv{\'e}n (whistler) branch, including strongly damped ion (electron) cyclotron waves.
However, such dispersive waves are harder to handle in analytic as well as in numerical approaches, since some of the assumptions used in standard QLT or magneto-hydrodynamic (MHD) theory no longer hold.
With the availability of increasingly powerful computers, numerical models have been established, which make use of a kinetic treatment of the plasma, thus surpassing restrictions of MHD -- such as the limitation to non-dispersive waves -- and being able to investigate different physical regimes.
For example, Particle-in-Cell (PiC) simulations, which are inherently self-consistent and able to model interactions of particles and electromagnetic fields on a micro-physical level, present one viable option to tackle the problem of dispersive waves and their interaction with charged particles \citep[e.g.][]{gary_2003,camporeale_2015}.
Also using PiC simulations, we showed in previous work that resonant interaction of protons and waves in the dispersive regime of the Alfv\'en branch is possible, although its application might be limited \citep{schreiner_2014_a}.

In this article we extend our study of wave-particle interaction, now focusing on the interaction of particles and right-handed, dispersive waves.
Such waves are observed in the interplanetary medium \citep[e.g.][and references therein]{droege_2000}, at interplanetary \citep[e.g.][]{aguilar-rodriguez_2011} and planetary bow shocks \citep[e.g.][]{fairfield_1974}, and in the Earth's foreshock region \citep[e.g.][]{palmroth_2015}.
Due to the occurrence of right-handed waves over a wide frequency range, resonant interactions with both protons and electrons are to be expected.
Being able to model and comprehend the scattering processes of particles and dispersive waves is a key requirement for understanding the complex dynamics of shocks and the turbulent interplanetary medium.
It is also an important step towards a more complete picture of diffusive shock acceleration, which is thought to accelerate particles up to very high energies and relativistic speeds.

Following our previous work, we present simulations showing resonant scattering of energetic electrons and right-handed, circularly polarized whistler waves.
We compare QLT predictions and results from both simple magnetostatic and full scale PiC simulations for various electron energies and different amplitudes of the whistler waves.
Parameter sets for the simulations are based on model parameters by \citet{vainio_2003} in order to resemble the conditions in the solar wind close to the Sun.

In Sect. \ref{sec:theory} we give a short description of resonant wave-particle scattering and the analytic model by \citet{lange_2013} which can be used to predict scattering amplitudes and which forms the basis for our analysis.
We then present our extension to this model in detail.
A brief overview of the numerical methods used for our simulations can be found in Sect. \ref{sec:numerics}, together with the setups and parameters of the individual simulations discussed later.

The results obtained from our PiC simulations are presented in Sects. \ref{sec:simulations_1} and \ref{sec:simulations_2}.
A further discussion of simulation results is then given in Sect. \ref{sec:discussion_diffusion}.
Finally, Sect. \ref{sec:closing_remarks} contains a short summary and conclusions.

\section{THEORY OF WAVE-PARTICLE INTERACTION}
\label{sec:theory}

The scattering of charged particles and plasma waves was first described as resonant interaction by \citet{jokipii_1966} using the so-called quasi-linear theory (QLT).
The basic idea of QLT is that field fluctuations $\delta \! F$ are small compared to a background field structure $F_0$: $(\delta \! F / F_0)^2 < 1$ \citep{schlickeiser_1994}.
With the absence of large scale electric fields in a plasma a magnetic field plays the role of the background structure and the particles are assumed to follow undisturbed gyro-orbits about the background field.
The QLT approach yields a condition for resonance \citep[e.g.][Chapter 12]{schlickeiser_2003}, which connects properties of a wave, such as its frequency $\omega$ and its wave vector $\boldsymbol{k}$, to the properties of a particle of species $\alpha$, such as its velocity $\boldsymbol{v}_\alpha$ and non-relativistic gyro frequency $\Omega_\alpha = q_\alpha \, B_0 / (m_\alpha \, c)$:
\begin{equation}
	\omega - k_\parallel \, v_\alpha \, \mu = n \, \frac{\Omega_\alpha}{\gamma_\alpha}.
	\label{resonance}
\end{equation}
Here, $k_\parallel$ and $v_\alpha \, \mu$ are the components of the wave vector and the particle's velocity parallel to the background magnetic field $\boldsymbol{B_0}$, where $\mu$ is the cosine of the pitch angle, i.e. the angle between $\boldsymbol{v}_\alpha$ and $\boldsymbol{B_0}$.
To allow for relativistic speeds, the particle's Lorentz factor $\gamma_\alpha$ is introduced to give the correct gyro frequency $\Omega_\alpha / \gamma_\alpha$.
Finally, $n$ is the order of the resonance, which will be set to $|n| = 1$ throughout this article.
With this choice of $n$ we limit ourselves to parallel propagating waves ($|\boldsymbol{k}| = |k_\parallel|$) with transverse polarization, which leads to simplifications both in the dispersion relation of the wave and in the theoretical approach used to describe pitch angle scattering.
Parallel or anti-parallel propagation is denoted by $k_\parallel > 0$ or $k_\parallel < 0$, respectively.

With given properties of a low frequency plasma wave and a characteristic speed of scattered particles, the resonance condition, Eq. (\ref{resonance}), yields a resonant pitch angle $\mu_\mathrm{res}$.
Particles propagating at an angle of $\mu_\mathrm{res}$ thus will be affected most efficiently by the scattering process, meaning that their pitch angle will change considerably.
Meanwhile, particles at different pitch angles will mostly bounce off of the magnetic field of the wave without a major change in their direction of motion.

\subsection{Magnetostatic Limit}
\label{sec:magnetostatic_frame}

For a more comfortable handling of wave-particle scattering theory, analytic descriptions are often derived in the so-called magnetostatic limit.
This limit can be seen as a transformation into the rest frame of the plasma wave, i.e. a frame of reference which moves with the phase speed of the wave $v_\mathrm{ph} = \omega / k$.
The key advantages of this transformation are that the electric field of the wave vanishes and the wave becomes a mere spatial modulation of the magnetic field with zero frequency.
Particles thus only interact with magnetic fields and their kinetic energy is conserved.
This can also be seen in the Fokker-Planck diffusion coefficients, where the only non-vanishing coefficient in a strict magnetostatic model is the pitch angle diffusion coefficient $D_{\mu\mu}$ \citep{schlickeiser_1994}.

In the following, primed symbols refer to quantities in the magnetostatic limit (i.e. the ``wave frame''), whereas unprimed symbols denote quantities in the rest frame of the plasma (the ``plasma frame'').
For example, the resonance condition (\ref{resonance}) in the wave frame becomes
\begin{equation}
	-k'_\parallel \, v'_\alpha \, \mu' = n \, \frac{\Omega'_\alpha}{\gamma'_\alpha}.
	\label{resonance_static}
\end{equation}

Based on the work of \citet{lee_1974}, \citet{lange_2013} have used the magnetostatic QLT approach to model the time evolution of the so-called scattering amplitude $\Delta\mu' (\mu',t')$, which describes the change of a particle's pitch angle $\Delta\mu'$ as a function of its pitch angle $\mu'$ and time $t'$.
We would like to stress that care has to be taken with the definitions of the pitch angle, $\mu'$, and the change thereof $\Delta\mu'$.
The change of the pitch angle is defined as the difference $\Delta\mu' := \mu'_t - \mu'_0$, where $\mu'_t := \mu'(t)$ and $\mu'_0 := \mu'(t=0)$ are the particles' pitch angles at the start and at the end of the interaction with the wave.
Although \citet{lange_2013} do not state this explicitly, their $\mu'$ which enters the equation for $\Delta\mu'$ presented below is neither $\mu'_t$ nor $\mu'_0$, as one might expect, but $\bar{\mu}' := (\mu'_t + \mu'_0) / 2$, i.e. the mean pitch angle cosine during the interaction.
The approximation $\bar{\mu}' \simeq \mu'_0$ which has been employed by \citet{lange_2013} might hold for $\Delta\mu' \ll \mu'_0$, but generally one should distinguish between $\mu'_0$, $\mu'_t$ and $\bar{\mu}'$.

As an analytic expression for the scattering amplitude (again only for $|n| = 1$) \citet{lange_2013} provide:
\begin{equation} 
	\Delta\mu'^{\pm}(\bar{\mu}',t',\Psi'^\pm) = \frac{\Omega'_\alpha}{\gamma'_\alpha} \frac{\delta \! B'}{B'_0} \sqrt{1-\bar{\mu}'^{2}} \cdot \frac{\cos{\left(\Psi'^\pm\right)} - \cos{\left((\pm k'_\parallel \, v'_\alpha \, \bar{\mu}' - \Omega'_\alpha / \gamma'_\alpha) \, t' + \Psi'^\pm\right)}}{\pm k'_\parallel \, v'_\alpha \, \bar{\mu}' - \Omega'_\alpha / \gamma'_\alpha},
	\label{qlt-deltamu}
\end{equation}
with the amplitude of the background magnetic field $B'_0$, the amplitude of the wave's magnetic field $\delta \! B'$ and the phase angle $\Psi'^\pm$ that maximizes $\Delta\mu'$:
\begin{equation} 
	\Psi'^\pm(\bar{\mu}',t') = \arctan{\left(\frac{\sin{\left((\pm k'_\parallel \, v'_\alpha \, \bar{\mu}' - \Omega'_\alpha / \gamma'_\alpha) \, t'\right)}}{1 - \cos{\left((\pm k'_\parallel \, v'_\alpha \, \bar{\mu}' - \Omega'_\alpha / \gamma'_\alpha) \, t'\right)}}\right)}.
	\label{qlt-psi}
\end{equation}
The two solutions of the above equations, denoted by superscript ($+$) and ($-$), describe the scattering of particles and waves with right-handed or left-handed polarization, respectively.
The full set of solutions is obtained by evaluating $\Delta\mu'^\pm (\bar{\mu}',t',\Psi'^\pm)$ and $\Delta\mu'^\pm (\bar{\mu}',t',\Psi'^\pm + \pi)$.
Equations (\ref{qlt-deltamu}) and (\ref{qlt-psi}) are derived under the assumption of test particles, i.e. particles which are affected by electromagnetic fields but which themselves do not act upon the fields.
The chosen theoretical framework could be called ``test particle QLT in the magnetostatic limit''.
For easier reference we will refer to this model as ``magnetostatic QLT'' in the following, although we do not want to imply that QLT is always restricted to the magnetostatic limit or to a test particle approach.

Note that the whole derivation of Eq. (\ref{qlt-deltamu}) does not make use of the random phase approximation which is otherwise typical for QLT.
The calculations of \citet{lange_2013} are based on Eq. (62) of \citet{lee_1974}, which describes the time derivative of the pitch angle cosine of a single particle with a defined phase angle.
Thus, Eq. (\ref{qlt-deltamu}) gives the scattering amplitude $\Delta\mu'$ of a particle with phase angle $\Psi'$ as a function of $\bar{\mu}'$.
Defining $\Psi'$ according to Eq. (\ref{qlt-psi}) means that a particle with the optimal phase angle in terms of scattering efficiency is chosen.

\subsection{Particle Scattering in the Plasma Frame}
\label{sec:plasma_frame}

Although the magnetostatic approximation is a valid approach in some physical regimes, it is only of limited use in other regimes.
If several plasma waves are to be considered, the magnetostatic approach is only valid if all waves propagate with the same phase speed.
This can only be achieved in the case of non-dispersive waves, such as Alfv\'en waves.
Dispersive waves with phase speeds depending on their frequency pose significant problems to standard magnetostatic theory.

Even though dispersive wave modes are harder to handle than Alfv\'en waves, theorists also explore the interaction of particles and dispersive waves.
For example, \citet{steinacker_1992} and \citet{vainio_2000} combine QLT and the full dispersion relations of waves in cold plasmas to derive Fokker-Planck coefficients and resonant frequencies of dispersive waves for protons and electrons.
However, for the concrete problem of scattering amplitudes, as described by equations (\ref{qlt-deltamu}) and (\ref{qlt-psi}), we are not aware of any analytic model that may be used to cover particle scattering off several dispersive waves.
We therefore present an approach to transform the equations above into the plasma frame.
The transformation can be performed for each of several waves individually.
The total scattering amplitude in the plasma frame is assumed to be the superposition of the scattering amplitudes resulting from the individual waves.

The idea of our approach is that equations (\ref{qlt-deltamu}) and (\ref{qlt-psi}) are valid in the rest frame of the wave.
Physical quantities measured in the plasma frame can then be transformed into the wave frame by the use of a simple Lorentz transformation, since the wave frame is boosted with the phase speed $v_\mathrm{ph} = \omega / k$ of the wave parallel or anti-parallel to the background magnetic field $\boldsymbol{B_0}$.
We examine the general case in which both the particle speed $v_\alpha$ and the phase speed $v_\mathrm{ph}$ can be relativistic.
This generalization leads to the occurrence of two Lorentz factors $\gamma_\alpha$ and $\gamma_\mathrm{w}$, where the former refers to the particles and the latter to the wave.

It is worth noting that the transformations used earlier by \citet{lange_2013} and \citet{schreiner_2014_a},
\begin{equation}
	\mu' = \mu - \frac{v_\mathrm{ph}}{v_\alpha}, ~~~ v'_\alpha = v_\alpha,
	\label{trans_old}
\end{equation}
are in fact never correct.
Although \citet{lange_2013} have shown that their transformations yield reasonable results in the case $v_\mathrm{ph} / v_\alpha \ll 1$, the range of $\mu'$ never covers the complete interval $-1 \leq \mu' \leq 1$.

In the following we give the correct transformations needed to express the primed quantities in equations (\ref{qlt-deltamu}) and (\ref{qlt-psi}) in the wave frame as functions of the unprimed quantities in the plasma frame:
\begin{align}
	t' &= t \, \gamma_\mathrm{w} \, \left(1 - \frac{v_\alpha \, \mu \, v_\mathrm{ph}}{c^2}\right),
	\label{trans_t}
	\\
	k'_\parallel &= k_\parallel / \gamma_\mathrm{w},
	\label{trans_k}
	\\
	\delta \! B' &= \delta \! B / \gamma_\mathrm{w},
	\label{trans_dB}
	\\
	\mu' &= \frac{v_\alpha \, \mu - v_\mathrm{ph}}{\sqrt{v^2_\alpha - 2 \, v_\alpha \, \mu \, v_\mathrm{ph} + v^2_\mathrm{ph} - v^2_\alpha \, v^2_\mathrm{ph} \, c^{-2} \, \big(1 - \mu^2\big)}},
	\label{trans_mu}
	\\
	v'_\alpha &= \frac{\sqrt{v^2_\alpha - 2 \, v_\alpha \, \mu \, v_\mathrm{ph} + v^2_\mathrm{ph} - v^2_\alpha \, v^2_\mathrm{ph} \, c^{-2} \, \big(1 - \mu^2\big)}}{1 - v_\alpha \, v_\mathrm{ph} \, c^{-2} \, \mu}.
	\label{trans_v}
\end{align}
Note that $B'_0 = B_0$ for a transformation between rest frames which are boosted parallel to $\boldsymbol{B_0}$ and thus $\Omega'_\alpha = \Omega_\alpha$.
The Lorentz factor $\gamma'_\alpha$ of the particles in the wave frame can then be computed using the expression for $v'_\alpha$.
We assume $v_\alpha$ to be constant, although this is not strictly the case as only $v'_\alpha$ is conserved in the wave frame.
However, under the assumption of low frequency waves with small phase speeds $v_\mathrm{ph} / c \ll 1$ the electric field $\delta \! E \sim \delta \! B \, v_\mathrm{ph} / c$ is much smaller than the magnetic field and the Fokker-Planck coefficients for momentum diffusion, $D_{\mu p}$ and $D_{pp}$, are also of order $v_\mathrm{ph} / c$ or $(v_\mathrm{ph} / c)^2$, respectively \citep{schlickeiser_1994}.
Thus, pitch angle diffusion can be assumed to be the fastest diffusion process.
Changes in particle energy, therefore, should be small and a constant particle speed may be assumed for a short enough period of time.

After the transformation of all relevant quantities in the plasma frame into the wave frame the scattering amplitude $\Delta\mu'^{\pm} (\bar{\mu}', t')$ can be computed according to equation (\ref{qlt-deltamu}).
However, this procedure only yields $\Delta\mu'(\bar{\mu}, t)$, i.e. the scattering amplitude in the wave frame as a function of the initial pitch angle and time in the plasma frame.
Thus, $\Delta\mu'$ itself has to be transformed back into the plasma frame.

The pitch angle cosine in the wave frame can be transformed into the plasma frame with a transformation similar to Eq. (\ref{trans_mu}):
\begin{align} 
		\mu = \frac{v'_\alpha \, \mu' + v_\mathrm{ph}}{\sqrt{v'^2_\alpha + 2 \, v'_\alpha \, \mu' \, v_\mathrm{ph} + v^2_\mathrm{ph} - v'^2_\alpha \, v^2_\mathrm{ph} \, c^{-2} \, \big(1 - \mu'^2\big)}},
	\label{backtrans_mu}
\end{align}
We can then compute $\Delta\mu(\bar{\mu},t)$, which can be expressed by the following equation:
\begin{align}
	\Delta\mu^{\pm} (\bar{\mu}, t) &= \mu_t - \mu_0
	\nonumber
	\\
	&= 2 \, (\mu_t - \bar{\mu})
	\nonumber
	\\
	&= 2 \, \big(\mu(\mu'_t) - \bar{\mu}\big)
	\nonumber
	\\
	&= 2 \, \Big(\mu\big(\Delta\mu'^{\pm}(\bar{\mu}, t) / 2 + \bar{\mu}'(\bar{\mu})\big) - \bar{\mu}\Big),
	\label{backtrans_deltamu}
\end{align}
where the application of equation (\ref{backtrans_mu}) to $\mu'_t = \Delta\mu'/2 + \bar{\mu}'$ yields $\mu_t$.

\begin{figure}[p]
	\centering
	\includegraphics[width=1.0\linewidth]{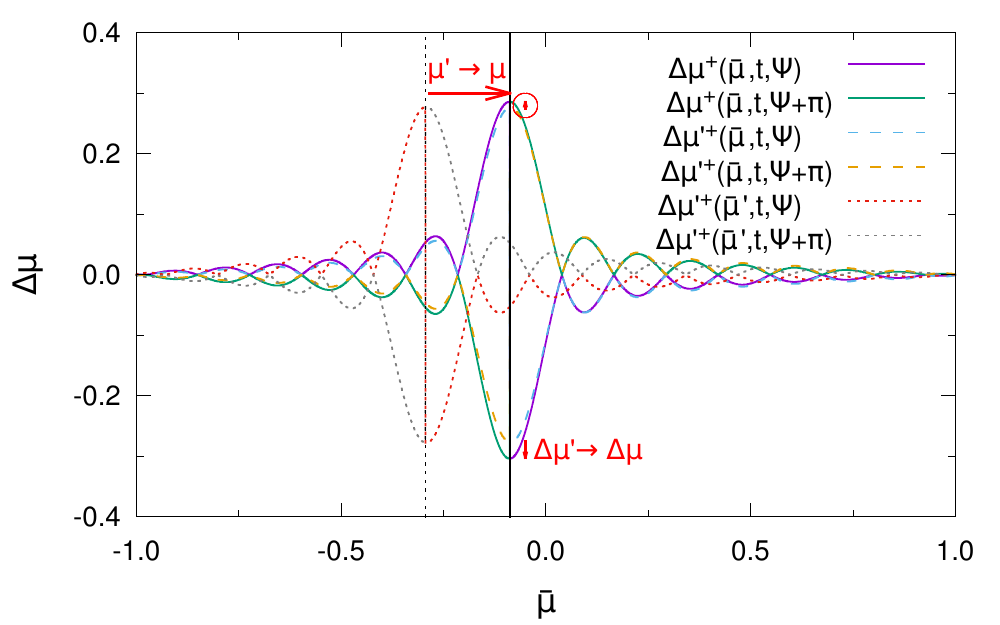} 
	\caption{
		Scattering amplitudes for the interaction of energetic electrons with a whistler wave propagating parallel to the background magnetic field.
		Dotted lines represent the scattering amplitude $\Delta\mu'$ as a function of $\bar{\mu}'$ in the wave frame, dashed lines are half-way through the transformations, representing $\Delta\mu'$ in the wave frame as a function of $\bar{\mu}$ in the plasma frame.
		Finally, the solid lines show $\Delta\mu$ as a function of $\bar{\mu}$ in the plasma frame and are the result of the full set of our transformations.
		Red arrows indicate the main effect of the transformation $\mu' \rightarrow \mu$ (one horizontal arrow) and $\Delta\mu' \rightarrow \Delta\mu$ (two vertical arrows), as labeled.
		Black vertical lines mark the positions of the resonance according to equations (\ref{resonance}) and (\ref{resonance_static}) in the plasma frame (solid line) and in the wave frame (dotted line).
	}
	\label{fig:scattering_amplitude_transformation}
\end{figure}

The transformations presented in this section might be cumbersome to carry out, but they are essentially straight forward.
However, one intrinsic problem arises from the last step, the back transformation of $\mu'_t$:
The final $\Delta\mu^{\pm} (\bar{\mu}, t)$ is not symmetric around $\Delta\mu = 0$, i.e. the scattering amplitudes are different for particles scattering to larger ($\Delta\mu > 0$) or smaller ($\Delta\mu < 0$) pitch angle cosines, as can be seen in Fig. \ref{fig:scattering_amplitude_transformation}.
This effect is not present in the wave frame (both when starting in the magnetostatic limit and when transforming plasma frame quantities into the wave frame) and is also not expected to occur in the plasma frame.
It is also not observed in self-consistent PiC simulations, as we will show later in Sect. \ref{sec:results_low_energy}, Fig. \ref{fig:scatter_plot_elec1}.

Figure \ref{fig:scattering_amplitude_transformation} shows a graphical representation of the transformations presented in this section.
The physical parameters that enter the equations used to produce the plot are taken from one of our simulations and represent electron scattering off a parallel propagating whistler wave.
The dotted lines are the results of equations (\ref{qlt-deltamu}) and (\ref{qlt-psi}) and depict the scattering amplitude in the wave frame.
Applying transformations (\ref{trans_t}) through (\ref{trans_v}) yields the dashed curves, which represent a mixture of both rest frames, since they depict the scattering amplitude $\Delta\mu'$ in the wave frame as a function of $\bar{\mu}$ in the plasma frame.
The red arrow labeled ``$\mu' \rightarrow \mu$'' indicates that these transformations mainly shift the QLT functions horizontally, i.e. only along the $\bar{\mu}$-axis.
Note that the amplitudes of the resonance and the smaller peaks remain constant.
After applying transformation (\ref{backtrans_deltamu}) the solid curves are obtained.
These curves represent $\Delta\mu$ as a function of $\bar{\mu}$ in the plasma frame.
The tiny, unlabeled vertical arrow at the top (inside the red circle) as well as the second vertical arrow labeled ``$\Delta\mu' \rightarrow \Delta\mu$'' at the bottom indicate that the final transformation only changes the scattering amplitude $\Delta\mu$ without shifting the curves further along the abscissa.
The different lengths of the arrows depict the asymmetry of the scattering amplitude which is caused by the transformation of $\Delta\mu$.

The general interpretation of the curves in Fig. \ref{fig:scattering_amplitude_transformation} is as follows:
Theory yields the maximum scattering amplitudes $\Delta\mu(\bar{\mu})$ for the mean pitch angle cosine $\bar{\mu}$, i.e. the pitch angle of a particle at a given $\bar{\mu}$ might change by any value between $-\Delta\mu(\bar{\mu})$ and $+\Delta\mu(\bar{\mu})$ depending on the particle's phase angle.
The largest changes in $\bar{\mu}$ (or $\mu_t$, since $\mu_0$ has a constant value) are possible at the resonance, i.e. at a mean pitch angle $\mu_\mathrm{res}$, as indicated by the dominant peaks in Fig. \ref{fig:scattering_amplitude_transformation}.
Far away from the resonance scattering is inefficient and most particles are hardly affected by the magnetic field of the wave.
This leads to so-called ballistic transport, which is sometimes defined as (scatter-) free propagation of particles \citep[e.g.][]{trotta_2011}.
However, some non-resonant or ``ballistic interactions'' \citep{lange_2013} are possible, as indicated by the smaller peaks in Fig. \ref{fig:scattering_amplitude_transformation}.
We will therefore refer to these smaller peaks as ``ballistic peaks''.

\subsection{Dispersion Relations}
\label{sec:dispersion_relations}

The waves relevant to our studies are parallel propagating, purely transverse waves with circular polarization.
In the cold plasma approximation, those waves are described by the dispersion relations \citep[e.g.][]{stix_1992}:
\begin{equation}
	|k^\pm_\parallel| = \frac{\omega}{c} \sqrt{1 - \frac{\omega_\mathrm{p}^2}{(\omega \pm \Omega_\mathrm{e}) (\omega \pm \Omega_\mathrm{p})}},
	\label{disp}
\end{equation}
with the plasma frequency $\omega_\mathrm{p}$ of an electron-proton plasma, the gyro frequencies of electrons and protons $\Omega_\mathrm{e}$ and $\Omega_\mathrm{p}$ and the speed of light $c$.
Note that the gyro frequency of the electrons $\Omega_\mathrm{e}$ has a negative sign.
The ($+$) and ($-$) solutions are valid for right- and left-handed waves, respectively.
These dispersion relations contain the whistler and Alfv\'en modes in their low frequency regimes.

\section{NUMERICAL APPROACH AND SIMULATION SETUP}
\label{sec:numerics}

\subsection{Code Overview}
\label{sec:code_overview}

To model dispersive waves of different kinds and to obtain a self-consistent description of electromagnetic fields and charged particles in the plasma, we employ a fully kinetic Particle-in-Cell approach \citep[see e.g.][]{hockney_1988}.
In particular, we use the PiC code \emph{ACRONYM} \citep{kilian_2011}, which employs an explicit second-order scheme and is fully relativistic, parallelized and three dimensional.
Although the PiC method might not be the most efficient numerical technique when dealing with proton effects, we still favor this approach for its versatility.
A more detailed discussion of advantages and drawbacks, as well as a direct comparison of PiC and MHD approaches to the specific problem of the interaction of protons and left-handed waves can be found in \citet[Sects. 3 and 6]{schreiner_2014_a}.
However, the PiC approach is well-suited for the study of electron scattering, since the time and length scales of electron interactions are closer to the scales of time step lengths and cell sizes in PiC simulations, thus reducing computing time compared to simulations in which proton interactions are studied.

The general setup for our simulations has also been characterized in previous works \citep[Sects. 3 and 3.2, respectively]{schreiner_2014_a,schreiner_2015} and will be outlined only briefly in this article.

In a thermal, magnetized electron-proton plasma one or more low-frequency waves are excited by choosing specific initial conditions for the electromagnetic fields and particle velocities.
This is done by solving the dispersion relation (\ref{disp}) for a specific wave number $k_\parallel$ to obtain the wave's frequency $\omega$.
Since the waves in question are circularly polarized, the phase relations between the different components of the electric and magnetic fields can be easily found \citep[][chapter 1-4, equation (42)]{stix_1992}.
In case of a background magnetic field in $z$-direction, the following condition holds for the electric field:
\begin{align}
	\frac{\imath \, E_x}{E_y} = \left\{
	\begin{array}{lr}
		+1 ~~~ \text{for right-handed waves},\\
		-1 ~~~ \text{for left-handed waves}.
	\end{array}
	\right.
	\label{polarization}
\end{align}
The field itself is described by a plane wave with frequency $\omega$ and wave number $k_\parallel$.
The magnetic field of the wave can be derived via Maxwell's equations, once the electric field is set -- or vice versa.
In the case of several waves the procedure can be repeated for each wave which yields the total electric and magnetic field as a superposition of the fields of the individual waves.

With electromagnetic fields deployed accordingly throughout the simulation box, an initial particle boost is performed.
The Lorentz force due to the wave's electromagnetic fields acts upon the particles and adds an ordered velocity component to the random thermal motion of each particle.
This velocity can be estimated by integrating the particles' equation of motion analytically in the cold plasma approximation, following the approach described by \citet[][chapter 1-2]{stix_1992}.

Finally, a non-thermal population of test particles (either electrons or protons) is added.
Those test particles are initialized with a mono-energetic velocity distribution and an isotropic angular distribution.
With the properties of the background plasma and the desired wave for wave-particle scattering already fixed, only a resonant pitch angle $\mu_\mathrm{res}$ has to be chosen to determine the speed of the test particles:
\begin{equation} 
	v_\alpha = \left|\frac{k_\parallel \, \omega \, |\mu_\mathrm{res}| \pm |\Omega_\alpha| \, \sqrt{k_\parallel^2 \, \mu_\mathrm{res}^{2} + (\Omega_\alpha^2 - \omega^2) \, c^{-2}}}{k_\parallel^2\,\mu_\mathrm{res}^{2} + \Omega_\alpha^2 \, c^{-2}}\right|.
	\label{vel}
\end{equation}

At the start of the simulation the background plasma will already be in a state, in which the fluctuations in the electromagnetic fields are dominated by the fields of the initially excited wave.
The test particles will be mostly unaffected by fluctuations created by thermal background particles and will interact predominantly with the magnetic field of the amplified wave, thus undergoing scattering processes.

To monitor the test particles the components of their velocity vectors and the components of the local magnetic field vector at the particles' positions are written to disk in certain output intervals.
With this method it is possible to track each particle's speed and pitch angle both towards the static background magnetic field and towards the local magnetic field at its position throughout the simulation.
However, if not denoted otherwise, we define ``the pitch angle'' as the angle between a particle's velocity vector and the static background magnetic field $\boldsymbol{B_0}$.
With the stored particle data it is specifically also possible to obtain the scattering amplitude $\Delta\mu(\bar{\mu},t)$.
To compute the scattering amplitude from the simulation data, the change of each particle's pitch angle between the current time step $t$ and the initial time step $t_0=0$ has to be calculated:
\begin{equation}
	\Delta\mu(t) = \mu_t - \mu_0.
	\label{deltamu}
\end{equation}
Binning and plotting $\Delta\mu$ over $\bar{\mu}$ for all test particles yields a scatter plot showing the scattering amplitude at a given time $t$.
For the plots shown in this article 512 bins for both $\Delta\mu$ and $\bar{\mu}$ have been chosen.

\subsection{Test Particle Code}
\label{sec:test_particle_code}

In addition to the PiC code described in the previous section we have also produced a magnetostatic test particle code.
This was done by modifying the existing PiC code: the electromagnetic field solver is switched off.
That means that the initial field configuration -- i.e. the electromagnetic fields of the excited wave -- will persist throughout the simulation.
This can be interpreted as being in the rest frame of the wave, where the phase speed of the wave is zero.
By transforming into the rest frame of the wave, its electric field vanishes.
Therefore, we do not initialize it at the beginning of the simulation.
The complete field setup only consists of the static background magnetic field $B_0$ and the periodic spatial magnetic field fluctuations $\delta \! B$ of the excited wave, which are produced as described in Sect.~\ref{sec:numerics}.

The relatively small number of numerical particles in a PiC simulation leads to noise in the electromagnetic fields.
It also produces a spectrum of physically allowed plasma waves.
These fluctuations cannot occur in the test particle code, because the background particles cannot interact with the electromagnetic fields due to the missing field solver.
Therefore, the thermal population of background particles is of no use to the simulation and can be left out completely.
What remains are only the mono-energetic test particles, which also do not act upon the magnetic fields, but which are still affected by them.

This procedure reduces our PiC code to a pure test particle pusher and yields a very simple picture of wave-particle interaction: static magnetic fields (i.e. being in the rest frame of the wave) and test particles, which are affected by the fields while the back reaction is neglected.
The simulations carried out with this kind of code will be called ``test particle'' or ``magnetostatic'' simulations in the article.

As an option for the test particle code a wave can be initialized with an electric field as well and the wave can be moved ``by hand''.
That means that the electromagnetic fields are updated each time step according to expected phase speed $v_\mathrm{ph} = \omega / k$ of a wave with wave number $k$ and frequency $\omega$.

\subsection{Simulation Setups}
\label{sec:simulation_setup}

The individual setups of the different simulations discussed in this article are relatively similar.
Only some parameters are subject to change, while basic parameters are always the same.
To save computing time all simulations have been performed using the 2d3v version of the \emph{ACRONYM} code, meaning that only two spatial dimensions are resolved, while velocity and field vectors are still three-dimensional.
Previous studies have shown that a two-dimensional setup is sufficient to recover the general characteristics of the wave-particle resonance, including the amplitudes of resonance peaks and the time evolution of the scatter plots.
Therefore, fully three-dimensional PiC simulations are not required at this stage of our studies.

Constant parameters are described in the text below, all other settings can be found in Table \ref{tab:phys_paras}.

\begin{table*}[p]
	\caption{Physical simulation parameters.}
	\label{tab:phys_paras}
	\centering
	\resizebox{\linewidth}{!}{
	\begin{tabular}{l c c c c c c c c c c c}
		\noalign{\smallskip}
		\hline
		\noalign{\smallskip}
		simulation (code) & $B_0$ & $k_\parallel$ & $\omega$ & $\delta \! B$ & $v_\mathrm{e}$ & $E_\mathrm{kin}$ & $\mu_\mathrm{res}$ & $\mu'_\mathrm{res}$ \\
		\noalign{\smallskip}
		& $(G)$ & $(|\Omega_\mathrm{e}| \, c^{-1})$ & $(|\Omega_\mathrm{e}|)$ & $(B_0)$ & $(c)$ & $(\mathrm{keV})$ & &\\
		\noalign{\smallskip}
		\hline
		\noalign{\smallskip}
		S1 (PiC) & $0.20$ & $50.6$ & $0.58$ & $0.02$ & $0.063$ & $1.0$ & $-0.13$ & $-0.30$\\
		\noalign{\smallskip}
		S2 (PiC) & $0.20$ & $50.6$ & $0.58$ & $0.02$ & $0.099$ & $2.5$ & $-0.082$ & $-0.20$\\
		\noalign{\smallskip}
		S3 (PiC) & $0.20$ & $50.6$ & $0.58$ & $0.02$ & $0.20$ & $10.5$ & $-0.039$ & $-0.10$\\
		\noalign{\smallskip}
		\hline
		\noalign{\smallskip}
		S4 (PiC) & $2.00$ & $5.06$ & $0.58$ & $0.02$ & $0.56$ & $105$ & $-0.088$ & $-0.28$\\
		\noalign{\smallskip}
		S5 (PiC) & $2.00$ & $5.06$ & $0.58$ & $0.02$ & $0.72$ & $226$ & $-0.031$ & $-0.19$\\
		\noalign{\smallskip}
		S6 (PiC) & $2.00$ & $5.06$ & $0.58$ & $0.02$ & $0.86$ & $495$ & $\phantom{-}0.017$ & $-0.12$\\
		\noalign{\smallskip}
		S7 (PiC) & $2.00$ & $5.06$ & $0.58$ & $0.02$ & $0.95$ & $1052$ & $\phantom{-}0.053$ & $-0.068$\\
		\noalign{\smallskip}
		\hline
		\noalign{\smallskip}
		S8 (PiC) & $2.00$ & $5.06$ & $0.58$ & $0.04$ & $0.56$ & $105$ & $-0.088$ & $-0.28$\\
		\noalign{\smallskip}
		S9 (PiC) & $2.00$ & $5.06$ & $0.58$ & $0.08$ & $0.56$ & $105$ & $-0.088$ & $-0.28$\\
		\noalign{\smallskip}
		S10 (PiC) & $2.00$ & $5.06$ & $0.58$ & $0.16$ & $0.56$ & $105$ & $-0.088$ & $-0.28$\\
		\noalign{\smallskip}
		S11 (PiC) & $2.00$ & $5.06$ & $0.58$ & $0.32$ & $0.56$ & $105$ & $-0.088$ & $-0.28$\\
		\noalign{\smallskip}
		\hline
		\noalign{\smallskip}
		S12 (PiC) & $2.00$ & $5.06,$ & $0.58$ & $0.02$ & $0.56$ & $105$ & $-0.088$ & $-0.28$\\
		\noalign{\smallskip}
		          &        & $-5.06$ & $0.58$ & $0.02$ &        &       & $\phantom{-}0.088$ & $\phantom{-}0.28$\\
		\noalign{\smallskip}
		S13 (PiC) & $2.00$ & $2.53,$ & $0.23,$ & $0.08,$ & $0.56$ & $105$ & $-0.40$ & $-0.54$\\
		\noalign{\smallskip}
		          &        & $5.06,$ & $0.58,$ & $0.04,$ &        &       & $-0.088$ & $-0.28$\\
		\noalign{\smallskip}
		          &        & $0.76,$ & $0.76,$ & $0.02,$ &        &       & $-0.017$ & $-0.19$\\
		\noalign{\smallskip}
		          &        & $0.85$ & $0.85$ & $0.01$ &        &       & $\phantom{-}0.0033$ & $-0.15$\\
		\noalign{\smallskip}
		\hline
		\noalign{\smallskip}
		S14 (test part.$^a$) & $2.00$ & $5.06$ & $0.58$ & $0.02$ & $0.56$ & $105$ & $-0.088$ & $-0.28$\\
		\noalign{\smallskip}
		S15 (test part.) & $2.00$ & $5.06$ & $0.58$ & $0.02$ & $0.56$ & $105$ & $-0.088$ & $-0.28$\\
		\noalign{\smallskip}
		S16 (test part.) & $2.00$ & $5.06,$ & $0.58$ & $0.02$ & $0.56$ & $105$ & $-0.088$ & $-0.28$\\
		\noalign{\smallskip}
		                 &        & $-5.06$ & $0.58$ & $0.02$ &        &       & $\phantom{-}0.088$ & $\phantom{-}0.28$\\
		\noalign{\smallskip}
		S17  (test part.) & $2.00$ & $2.53,$ & $0.23,$ & $0.08,$ & $0.56$ & $105$ & $-0.40$ & $-0.54$\\
		\noalign{\smallskip}
		                 &        & $5.06,$ & $0.58,$ & $0.04,$ &        &       & $-0.088$ & $-0.28$\\
		\noalign{\smallskip}
		                 &        & $0.76,$ & $0.76,$ & $0.02,$ &        &       & $-0.017$ & $-0.19$\\
		\noalign{\smallskip}
		                 &        & $0.85$ & $0.85$ & $0.01$ &        &       & $\phantom{-}0.0033$ & $-0.15$\\
		\noalign{\smallskip}
		\hline
	\end{tabular}
	\tablenotetext{a}{Test particle simulation with moving wave.}
	}
\end{table*}

All simulations (except for S13 and S17) share the same numerical parameters, such as the numbers of cells in the directions parallel and perpendicular to the background magnetic field \mbox{$N_\parallel = 4096$} and \mbox{$N_\perp = 128$}, cell size \mbox{$\Delta x = 1.29 \cdot 10^{-3}\,c \,\omega_\mathrm{p}^{-1}$} and time step length \mbox{$\Delta t = 7.46 \cdot 10^{-4} \, \omega_\mathrm{p}^{-1}$}.
The box size corresponds to the wave length of the amplified whistler wave in parallel direction and twice the gyro radius of thermal electrons in the perpendicular direction.
By using the natural ratio of proton to electron mass the proton dynamics becomes negligible on the simulated time scales, so the proton gyration does not have to be resolved.
The number of test electrons is \mbox{$N_\mathrm{e} = 1.0 \cdot 10^6$} per simulation.

Simulations S13 and S17 are twice as long ($N_\parallel = 8192$) and host twice as many test electrons (\mbox{$N_\mathrm{e} = 2.1 \cdot 10^6$}).

The physical parameters for the simulations are based on model parameters by \citet{vainio_2003} resembling the (slow) solar wind at a distance of 2.6 solar radii.
To characterize the plasma we choose the field strength of the static background magnetic field $B^\mathrm{model}_0 = 0.20 \, \mathrm{G}$, the plasma frequency $\omega^\mathrm{model}_\mathrm{p} = 6.4 \cdot 10^7 \, \mathrm{rad} \, \mathrm{s}^{-1}$ and the temperature $T^\mathrm{model} = 2.0 \cdot 10^6 \, \mathrm{K}$ according to the model.
However, we adapt these model parameters in order to optimize our simulation in terms of computational effort.
Therefore, we employ a plasma frequency $\omega_\mathrm{p} = 1.5 \cdot 10^8 \, \mathrm{rad} \, \mathrm{s}^{-1}$ and a temperature $T = 2.0 \cdot 10^4 \, \mathrm{K}$ in all simulations.
The adjustment of the plasma frequency yields a more convenient cell size and time step length, whereas a reduced temperature shifts the onset of cyclotron damping of the parallel propagating waves to higher frequencies, thus allowing for the excitation of waves with shorter wave lengths.
As a consequence, the length of the simulation box can also be reduced.
For simulations of low energy test electrons (S1, S2, S3 and S14) we keep the magnetic field from the model, but for all other simulations we increase $B_0$ by a factor of ten (see Table \ref{tab:phys_paras}).

In terms of the electron gyro frequency and the thermal speed of the background electrons, the simulation parameters described above translate to \mbox{$|\Omega_\mathrm{e}| = 2.34 \cdot 10^{-2} \, \omega_\mathrm{p,e}$} (S1, S2, S3 and S14) or \mbox{$|\Omega_\mathrm{e}| = 2.34 \cdot 10^{-1} \, \omega_\mathrm{p,e}$} (other simulations) and \mbox{$v_\mathrm{th,e} = 1.83 \cdot 10^{-3} \, c$}.

One or more right-handed waves with parallel wave numbers $k_\parallel$ and frequencies $\omega$ are chosen and initialized in each simulation.
The magnetic field amplitude $\delta \! B$ of each amplified wave is set to a fraction of the background magnetic field strength.
The test electrons are initialized as mono-energetic particle populations with a kinetic energy $E_\mathrm{kin}$ and speed $v_\mathrm{e}$.
Solving the resonance condition in the plasma or in the wave frame, Eq. (\ref{resonance}) or Eq. (\ref{resonance_static}), for the resonant pitch angle in the respective rest frame yields $\mu_\mathrm{res}$ or $\mu'_\mathrm{res}$.
These parameters are given in Table \ref{tab:phys_paras} for each simulation.

Table \ref{tab:phys_paras} also contains information about the nature of the simulation, i.e. if it is a full PiC simulation or a test particle simulation (see Sect. \ref{sec:test_particle_code}).

For clarity the simulations in Table \ref{tab:phys_paras} are organized in blocks, separated by horizontal lines.
The first block contains simulations S1 through S3 (low energy test electrons), which are first described in Sect. \ref{sec:results_low_energy}.
Simulations S4 through S7 (high energy test electrons) are first analyzed in Sect. \ref{sec:results_high_energy}.
The third block (S8 through S11) lists simulations with changing amplitude of the wave, as discussed in Sect. \ref{sec:results_amplitude}.
In the two simulations S12 and S13 several waves are initialized.
Results from these simulations are presented in Sects. \ref{sec:results_two_waves} and \ref{sec:results_four_waves}.
Finally, the last block simply contains all simulations carried out with the test particle code.
Some of these simulations are first discussed in Sect. \ref{sec:results_magnetostatic}, others accompany their PiC counterparts in Sects. \ref{sec:results_two_waves} and \ref{sec:results_four_waves}.

\subsection{Note on the Use of Single Waves for Wave-Particle Scattering}
\label{sec:note_on_single_waves}

In order to derive a meaningful theory of pitch angle diffusion in the framework of QLT the so-called random phase approximation has to be made \citep[e.g.][chapter 12]{schlickeiser_2003}.
This means that the phase relation between particle and wave has to be randomized.
One can achieve the randomization in two ways: Either a large number of waves with different phases or a large number of particles with randomly chosen phases with respect to their gyration about the ordered magnetic field (``gyrophase'') is required.
Usually a large number of waves is assumed, which is consistent with the idea of turbulence and a cascade of waves with various wave lengths and frequencies.

By default, we study the interaction of particles and a single wave in our simulations.
However, we employ a large sample of test particles, which are initialized with random directions of motion and thus random gyrophases.
Therefore, the random phase approximation is valid when looking at the whole ensemble of test particles.

Additionally, as we are not considering whole spectra of waves, but only waves with a clearly defined wave number and frequency, the many-waves approach would fail:
Assume that several waves with the same wave length and frequency, but different phase angles and amplitudes are initialized.
The combined electromagnetic field would be the superposition of the electromagnetic fields of each single wave.
Each component $F$ of the electric or magnetic field could then be written as
\begin{equation}
	F(k,\omega) = \sum\limits_n \, A_n \, \sin\left(\vec{k}\cdot\vec{r} - \omega \, t + \phi_n\right) = A \, \sin\left(\vec{k}\cdot\vec{r} - \omega \, t + \xi\right),
	\label{combined_field}
\end{equation}
where $A_n$ and $\phi_n$ are the amplitudes and phase angles of the individual waves and $A$ and $\xi$ are the amplitude and phase of the resulting superposition.
Thus, the superposition of the waves can again be seen as a single wave.

\section{SIMULATIONS I: VARIATION OF THE PARTICLE ENERGY}
\label{sec:simulations_1}

In a first series of simulations we probe the effect of the electron energy on their scattering behavior.
To keep the variations in the setups for the single simulations at a minimum, we employ a constant size of the simulations box (both in numerical grid cells and in physical length scales) and the same physical parameters for the background plasma in all simulations (see Sect. \ref{sec:simulation_setup}).
However, particles with different energies prefer waves in different regimes of frequencies to resonate with and thus it is convenient to split our study into two subsets of simulations:
A first set of three simulations (S1 through S3) with low-energy test electrons (\mbox{$E_\mathrm{kin} \sim1 - 10\,$keV}) and a second set of four simulations (S4 through S7) with high-energy test electrons (\mbox{$E_\mathrm{kin} \sim 100 - 1000\,$keV}).

\subsection{Low-Energy Electrons}
\label{sec:results_low_energy}

\begin{figure}[p]
	\centering
	\includegraphics[width=0.64\linewidth]{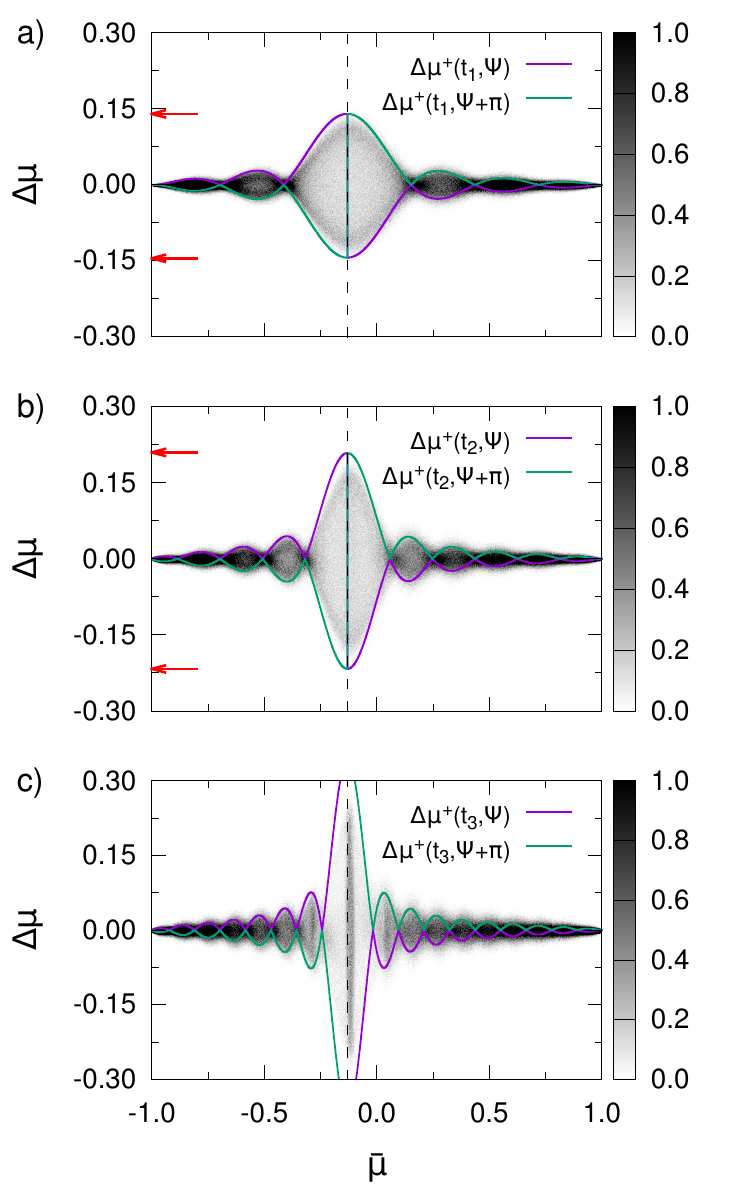}
	\caption{
		Scatter plots showing test electron data from simulation S1 and the theoretical scattering amplitude $\Delta\mu^+ \, (\bar{\mu},t,\Psi^+)$ (solid lines) according to equation (\ref{backtrans_deltamu}) at three points in time \mbox{$t_1 \cdot |\Omega_\mathrm{e}| = 7.0$} (panel a), \mbox{$t_2 \cdot |\Omega_\mathrm{e}| = 10.5$} (panel b) and \mbox{$t_3 \cdot |\Omega_\mathrm{e}| = 17.5$} (panel c).
		Dashed lines indicate the position of the resonance according to the resonance condition, equation (\ref{resonance}).
		Color coding represents test electron density in $\bar{\mu}$-$\Delta\mu$ space, normalized to $10^{-4} \cdot N_\mathrm{e}$, where $N_\mathrm{e}$ is the total number of test electrons per simulation.
		Red arrows in panels a) and b) mark the hight of the resonance peak for positive and negative $\Delta\mu$.
	}
	\label{fig:scatter_plot_elec1}
\end{figure}

We first analyze simulation S1, which includes test electrons of the lowest energy.
Figure \ref{fig:scatter_plot_elec1} illustrates the time evolution during $10^6$ time steps in the simulation and shows the growth and saturation of the resonance peak.
During the initial growth phase of the resonance, simulation results and magnetostatic QLT predictions match well (Fig. \ref{fig:scatter_plot_elec1}, panel a) except for the asymmetry of the predicted $\Delta\mu$ which was mentioned in Sect. \ref{sec:plasma_frame}.
At later time steps analytic predictions overestimate the amplitude of the resonance and the asymmetry increases (see red arrows indicating peak height in panels a and b).
In this case, theoretical prediction and simulation diverge already while the resonance peak is still growing (panel b) and differences can be clearly seen at the end of the simulation (panel c).
Looking at the simulation data (color coded particle density in $\bar{\mu}$-$\Delta\mu$-space) panel c) shows a substructure inside the resonance peak.
These structures are observed to grow in amplitude and broaden, eventually replacing the previous resonant structure which fades away as particles are scattered away from the resonance.

In the simulation it is observed that the resonance peak reaches a stable amplitude, suggesting that the maximum change of the pitch angle per time interval is limited.
Magnetostatic QLT predictions from equation (\ref{qlt-deltamu}), however, develop a delta-shaped peak with further increasing amplitude as time advances.
The delta-shape of the resonance is, of course, unrealistic and the predicted amplitude is also not expected to occur in reality.
Besides the amplitude of the resonance, the analytic approximation matches very well with the contours of the electron data from the simulation, as can be seen in Fig. \ref{fig:scatter_plot_elec1}.
To obtain the theoretical curves the phase angle $\Psi$ which maximizes the scattering amplitude has been chosen in equation (\ref{qlt-psi}).
However, with the particles interacting at different positions along the wave not every particle ``sees'' the optimal phase for maximum scattering.
Thus, in the scatter plot the area inside the analytical curves is also populated with particles.

\begin{figure}[p]
	\centering
	\includegraphics[width=0.7\linewidth]{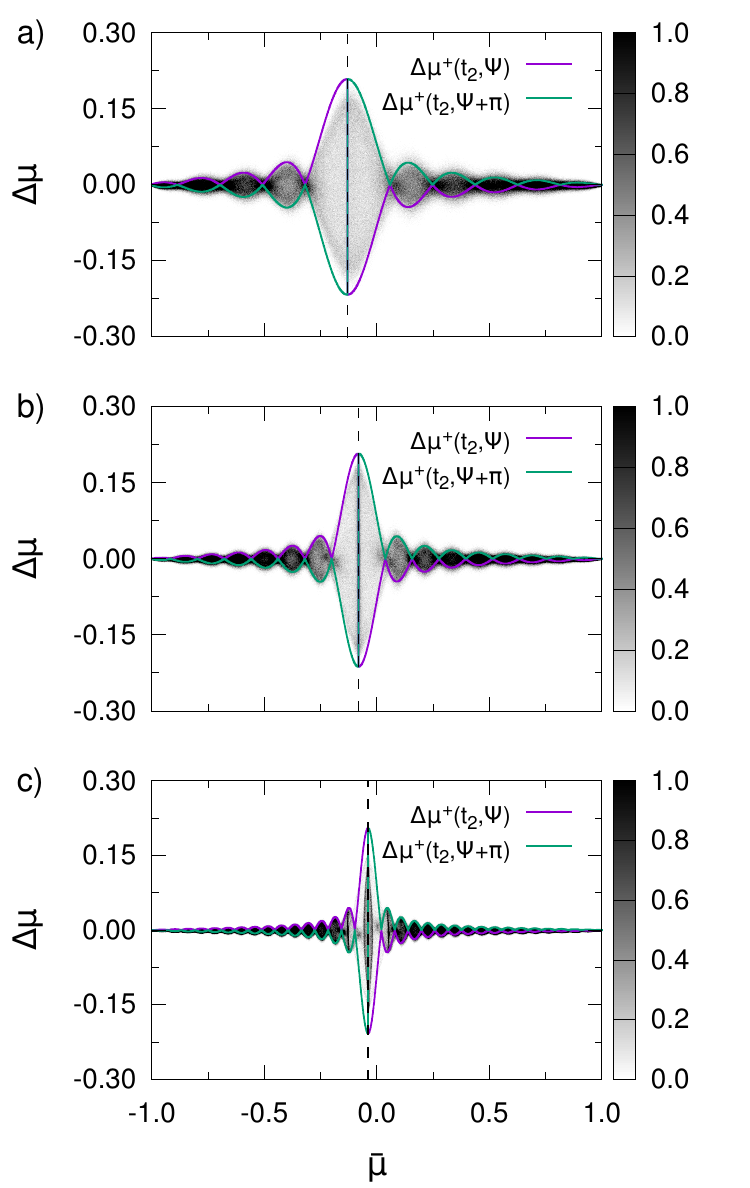}
	\caption{
		Scatter plots of test electrons with different kinetic energies at time \mbox{$t_2 \cdot |\Omega_\mathrm{e}| = 10.5$}.
		Panels a), b) and c) show data from simulations S1, S2 and S3, respectively, with panel a) here being the the same as Fig. \ref{fig:scatter_plot_elec1} b).
		Solid and dashed lines as well as color coding are the same as in Fig. \ref{fig:scatter_plot_elec1}.
	}
	\label{fig:scatter_plot_elec2}
\end{figure}

To examine the effect of particle energy on the scattering behavior we present results from the first set of simulations, S1, S2 and S3, in Fig. \ref{fig:scatter_plot_elec2}.
The three panels show scatter plots from the three simulations at the same point in time.
As seen before in Fig. \ref{fig:scatter_plot_elec1} the analytic approximation describes the data well, except for the hight of the resonance peak which is overestimated by theoretical predictions.
It can be seen that the number of side peaks from ballistic scattering increases with particle energy from panel a) to c) and that the resonance peak becomes narrower.
Note that the narrowing of the resonance is an effect of the chosen abscissa, which displays $\bar{\mu}$, whereas the resonances in all three panels would have the same width if the parallel velocity $v_\mathrm{e} \, \bar{\mu}$ was chosen as the abscissa.
However, the evolution of the resonances in the three simulations still differs, as can be seen when looking at the substructure of the resonance peak.
Whereas no substructure can be seen in panel a), a straight vertical line at $\mu_\mathrm{res}$ can be made out in panel b) and a wider structure is present in panel c).
This suggests that the wave-particle resonance of faster, more energetic particles evolves differently or maybe faster than for low energy particles.

\subsection{High-Energy Electrons}
\label{sec:results_high_energy}

The second set of simulations, S4 through S7, includes test electrons with higher energies ($105 ~ \mathrm{keV}$ to $1052 ~ \mathrm{keV}$) and can be analyzed in the same way as the first one.
Due to the increased background magnetic field $B_0$ (see Table \ref{tab:phys_paras} in Sect. \ref{sec:simulation_setup}), the frequency of the wave $\omega$ and the electron cyclotron frequency $\Omega_\mathrm{e}$ are also increased by a factor of $10$, compared to the first set of simulations (S1 through S3).
Therefore, the evolution of the resonance can be observed in a shorter physical time span (i.e. less numerical time steps of a given length).
However, if time is measured in units of the cyclotron frequency, time scales are comparable between the two series of simulations.

\begin{figure*}[p]
	\centering
	\includegraphics[width=1.\linewidth]{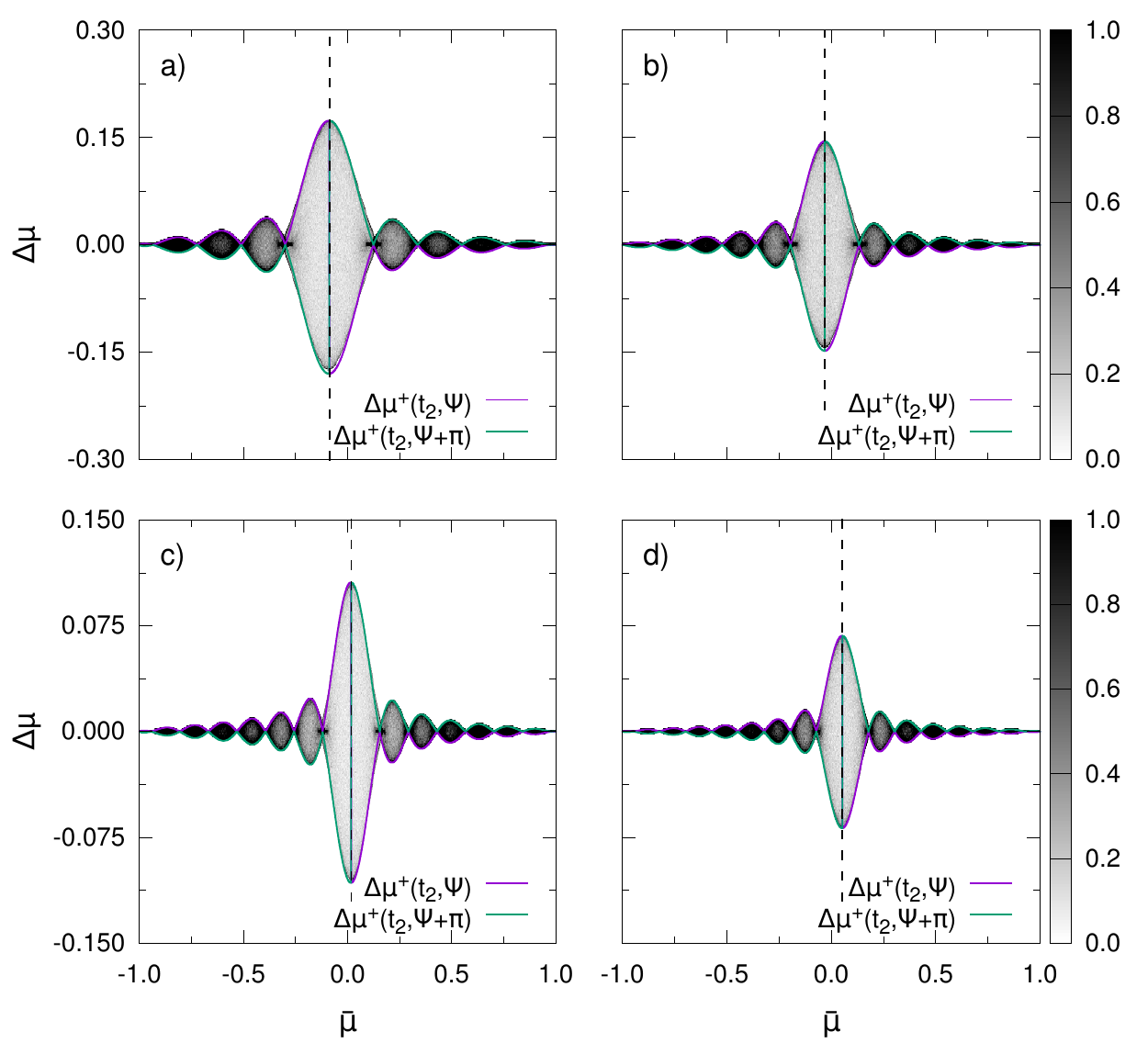}
	\caption{
		Scatter plots showing test electron data from simulations S4 (panel a), S5 (panel b), S6 (panel c) and S7 (panel d) for time \mbox{$t_2 \cdot \Omega_\mathrm{e} = 10.5$}.
		Solid and dashed lines as well as color coding are the same as in Fig. \ref{fig:scatter_plot_elec1}.
	}
	\label{fig:scatter_plot_elec3}
\end{figure*}

We show simulation data from the second set of simulations in Fig. \ref{fig:scatter_plot_elec3} at the same normalized time \mbox{$t_2 \cdot \Omega_\mathrm{e} = 10.5$}, which was already used for Fig. \ref{fig:scatter_plot_elec2} in Sect. \ref{sec:results_low_energy}.
Comparison of Fig. \ref{fig:scatter_plot_elec3} and Fig. \ref{fig:scatter_plot_elec2} reveals that the resonance peaks are at comparable stages of their evolution for both sets of simulations.

Within Fig. \ref{fig:scatter_plot_elec3} the four panels a) through d) refer to the four simulations S4 through S7.
Test electron energy thus increases from \mbox{$E_\mathrm{kin,e} = 105 \, \mathrm{keV}$} (panel a) to \mbox{$E_\mathrm{kin,e} = 1052 \, \mathrm{keV}$} (panel d), as stated in Sect. \ref{sec:simulation_setup}, Table \ref{tab:phys_paras}.
As already found for the low-energy electrons in Sect. \ref{sec:results_low_energy}, the number of side peaks increases and the resonance peak becomes narrower (as a result of the chosen abscissa) with increasing particle energy.
Shape and position of the resonance peaks are in accordance with magnetostatic QLT for all simulations.

Contrary to previous test cases of low-energy electrons, the Lorentz factor of the test particles increases significantly for the high-energy electrons.
This has a direct effect on the scattering amplitude, which decreases with increasing particle energy.
The amplitude of the resonance peak \mbox{$\Delta\mu_\mathrm{res} := \Delta\mu(\mu_\mathrm{res})$} is roughly inversely proportional to the Lorentz factor $\gamma_\mathrm{e}$ of the test electrons, as might be expected from equation~(\ref{qlt-deltamu}).
A comparison of the Lorentz factors $\gamma_\mathrm{e}$ and peak amplitudes $\Delta\mu_\mathrm{res}$ of the different simulations can be found in Table \ref{tab:scattering_amplitudes}, where an additional index $j$ denotes the number of the simulation.

\begin{table}[h]
	\caption[]{Comparison of scattering amplitudes and Lorentz factors.}
	\label{tab:scattering_amplitudes}
	\centering
	\begin{tabular}{c c c}
		\noalign{\smallskip}
		\hline
		\noalign{\smallskip}
		$j$ & $\Delta\mu_{\mathrm{res},j} \, / \, \Delta\mu_{\mathrm{res},4}$ & $\gamma_{\mathrm{e},4} \, / \, \gamma_{\mathrm{e},j}$\\
		\noalign{\smallskip}
		\hline
		\noalign{\smallskip}
		5 & $0.85$ & $0.84$\\
		\noalign{\smallskip}
		6 & $0.63$ & $0.61$\\
		\noalign{\smallskip}
		7 & $0.41$ & $0.40$\\
		\noalign{\smallskip}
		\hline
	\end{tabular}
\end{table}

\subsection{Magnetostatic Simulations of High-Energy Electrons}
\label{sec:results_magnetostatic}

\begin{figure*}[p]
	\centering
	\includegraphics[width=0.64\linewidth]{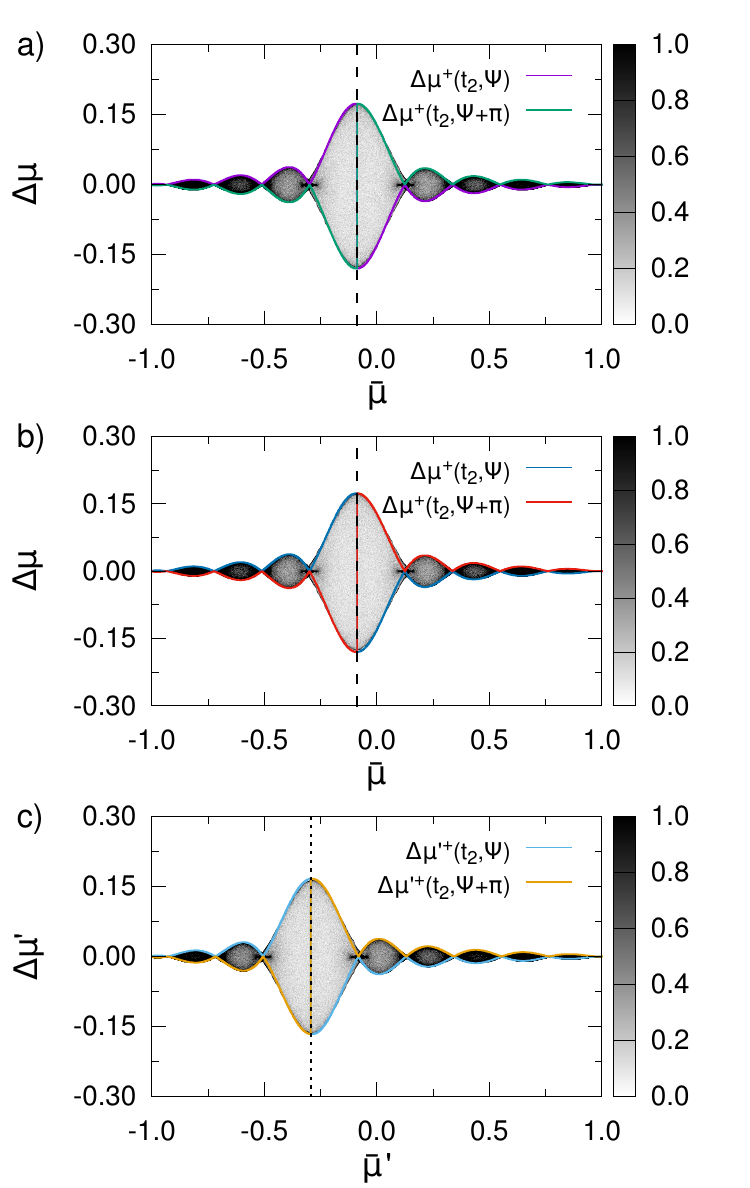}
	\caption{
		Scatter plots showing test electron data from full PiC simulations in the rest frame of the plasma (S4, panel a) and test particle simulations with a moving wave (S14, panel b) as well as in the rest frame of the wave (S15, panel c).
		The time step shown is \mbox{$t_2 \cdot |\Omega_\mathrm{e}| = 10.5$}.
		Solid lines represent magnetostatic QLT predictions in the plasma (panels a and b) and in the wave frame (panel c), dashed (dotted) lines mark the resonant pitch angles $\mu_\mathrm{res}$ ($\mu'_\mathrm{res}$) in the plasma (wave) frame.
		Color coding of the simulation data is the same as in Fig. \ref{fig:scatter_plot_elec1}.
	}
	\label{fig:scatter_plot_static}
\end{figure*}

To further validate our results we perform simulations in the magnetostatic limit using the test particle code described in Sect. \ref{sec:test_particle_code}.
The static magnetic field of the wave in the test particle simulations mimics the situation in the wave's rest frame and we expect perfect agreement of simulation results and theory.
A comparison of results from magnetostatic and full PiC simulations may reveal whether any additional effects may occur in the more realistic PiC simulations which are not captured by the magnetostatic approximation.

Figure \ref{fig:scatter_plot_static} shows the results of a full PiC simulations (S4, panel a) and two equivalent simulations which have been performed with the test particle code (see Sect. \ref{sec:test_particle_code}).
One of these simulations (S14, panel b) includes a wave which is artificially pushed forward in every time step, thus imitating a propagating wave.
The other test particle simulation (S15, panel c) incorporates the magnetostatic approximation, i.e. the magnetic field fluctuations are static.

Resonant scattering of electrons and the magnetic fields of the wave is also reproduced in the test particle simulations.
For the theoretical modeling the transformed equations leading to $\Delta\mu(\bar{\mu})$, Eq. (\ref{backtrans_deltamu}), in the plasma frame have to be used for S4 and S14 and the original equations of \citet{lange_2013}, Eqs. (\ref{qlt-deltamu}) and (\ref{qlt-psi}), have to be used for S15.
Numerical results and magnetostatic QLT predictions match perfectly for all three simulations.

The position of the resonance is identical in the PiC simulation and the test particle simulation with a moving wave, whereas the position is shifted in the magnetostatic test particle simulation, as illustrated in Fig. \ref{fig:scattering_amplitude_transformation} in Sect. \ref{sec:plasma_frame}.
This is, of course, an effect of the different rest frames: Whereas the resonant pitch angle $\mu_\mathrm{res}$ in the plasma frame is given by equation (\ref{resonance}), the corresponding $\mu'_\mathrm{res}$ in the wave frame is given by equation (\ref{resonance_static}).
Table~\ref{tab:phys_paras} gives the expected resonant pitch angles $\mu_\mathrm{res}$ and $\mu'_\mathrm{res}$, which are in agreement with the numerical results.
Besides the expected shift of the resonances no qualitative differences between the results of the three types of simulations can be found.
Although additional field fluctuations and electric fields are present in the PiC simulations, it seems that the scattering behavior of the energetic test electrons is still well-described by the magnetostatic approximation.

\begin{figure}[p] 
	\centering
	\includegraphics[width=0.75\linewidth]{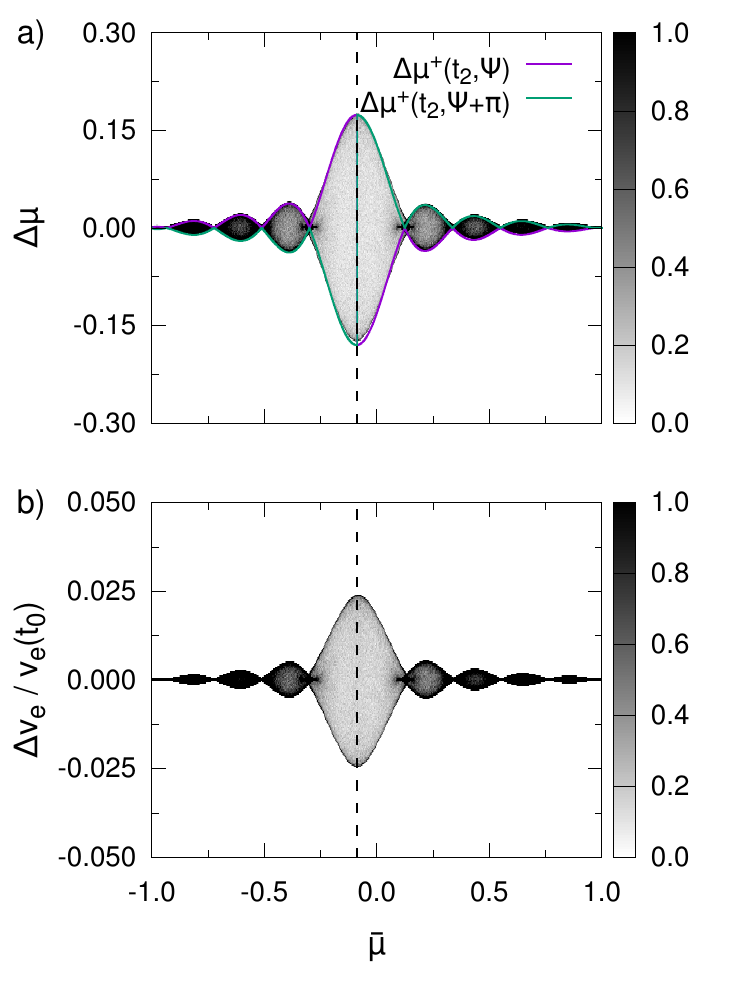} 
	\caption{
		Scatter plots showing both pitch angle scattering (panel a) and velocity scattering (panel b) as functions of the mean pitch angle $\bar{\mu}$.
		Data is taken from simulation S4, panel a) is the same as Fig. \ref{fig:scatter_plot_elec3} a).
		The change in particle velocity $\Delta \, v_\mathrm{e} = v_{\mathrm{e},2} - v_{\mathrm{e},0}$ compares the particles' speeds at the beginning of the simulation, $t_0$, and a later point in time, $t_2 \cdot |\Omega_\mathrm{e}| = 10.5$.
		The position of the resonance is marked by the dashed lines in both panels.
		Color coded simulation data is scaled as in Fig. \ref{fig:scatter_plot_elec1}.
	}
	\label{fig:scatter_plot_speed}
\end{figure}

As a further test we study the change of particle energy over time.
In the wave's rest frame the interaction of the particles and the wave are elastic, i.e. energy conserving.
In the rest frame of the plasma, however, the energy of the particles may increase or decrease depending on the relative propagation of particle and wave (``head-on'' or ``overtaking collision'', as suggested in the model of \citet{fermi_1949}).

We plot $(v_{\mathrm{e},t} - v_{\mathrm{e},0}) / v_{\mathrm{e},0}$ over the mean pitch angle $\bar{\mu}$, where $v_{\mathrm{e},t}$ is a particle's current speed at time $t$ and $v_{\mathrm{e},0}$ is its speed at the beginning of the simulation.
Figure \ref{fig:scatter_plot_speed} shows scatter plots of pitch angle scattering (panel a) and velocity scattering (panel b) for one point in time during simulation S4.
It can be seen that the particles' speeds change over time and in a fashion which is very similar to their change of pitch angle.
This behavior can be expected, as the following examination of energy and momentum changes in the plasma and the wave frame will show.
The energy $E'_\mathrm{e}$ of an electron in the wave frame can be expressed by its energy $E_\mathrm{e}$ and the parallel component $p_{\parallel \, \mathrm{e}}$ of the momentum vector in the plasma frame:
\begin{equation}
	E'_\mathrm{e} = \gamma_\mathrm{w} \, E_\mathrm{e} - \gamma_\mathrm{w} \, v_\mathrm{ph} \, p_{\parallel \, \mathrm{e}},
	\label{energy_wave_frame}
\end{equation}
where $v_\mathrm{ph}$ and $\gamma_\mathrm{w}$ are the phase speed and Lorentz factor of the wave.
Using the above equation and assuming $\Delta E'_\mathrm{e} = 0$ the condition
\begin{equation}
	\Delta E_\mathrm{e} = v_\mathrm{ph} \, \Delta p_{\parallel \, \mathrm{e}}
	\label{delta_energy1}
\end{equation}
is obtained.
The right hand side can be rewritten using $\Delta p_{\parallel \, \mathrm{e}} = \gamma_\mathrm{w} \, \Delta p'_{\parallel \, \mathrm{e}}$:
\begin{equation}
	\Delta E_\mathrm{e} = \gamma_\mathrm{w} \, v_\mathrm{ph} \, p' \, \Delta \mu',
	\label{delta_energy2}
\end{equation}
where $\Delta p'_{\parallel \, \mathrm{e}} = p' \, \Delta \mu'$ has been inserted.
This clearly predicts that momentum scattering and pitch angle scattering occur at the same positions in $\mu$-space.

In the example shown in Fig.~\ref{fig:scatter_plot_speed}, particles might change their speed by up to 2.5 percent of their initial speed.
The time scale at which particle energy changes seems to be large enough so that pitch angle scattering can be assumed to be the fastest diffusion process.
This was one of the prerequisites of the model presented in Sect. \ref{sec:theory}.

In the magnetostatic simulations no change of the particles' speeds is observed, as would be expected from the setup of these simulations.

\section{SIMULATIONS II: VARIATION OF THE WAVE AMPLITUDE}
\label{sec:simulations_2}

Since QLT assumes that the wave's amplitude $\delta \! B$ is small compared to the background magnetic field $B_0$, it is interesting to test the influence of different $\delta \! B$ on the evolution of wave-particle scattering.
Using the parameters from simulation S4 ($100 \, \mathrm{keV}$ electrons, see Sect. \ref{sec:results_high_energy}), we set up a third set of simulations, S8 through S11, in which we vary the amplitude of the wave's magnetic field.
Additionally, we test the effect of two counter-propagating waves with the same wave number $|k_\parallel|$, frequency $\omega$ and amplitude $\delta \! B$ (simulation S12).
A further simulation, S13, is set up which contains four waves with different amplitudes, wave numbers and frequencies.
The latter two simulations each have a magnetostatic counterpart, S16 and S17, carried out with the test particle code.
The relevant parameters for simulations S8 through S13, as well as S16 and S17 are given in Table \ref{tab:phys_paras} in Sect. \ref{sec:simulation_setup}.

\subsection{Effect of the Amplitude of a Single Wave}
\label{sec:results_amplitude}

\begin{figure*}[p] 
	\centering
	\includegraphics[width=1.\linewidth]{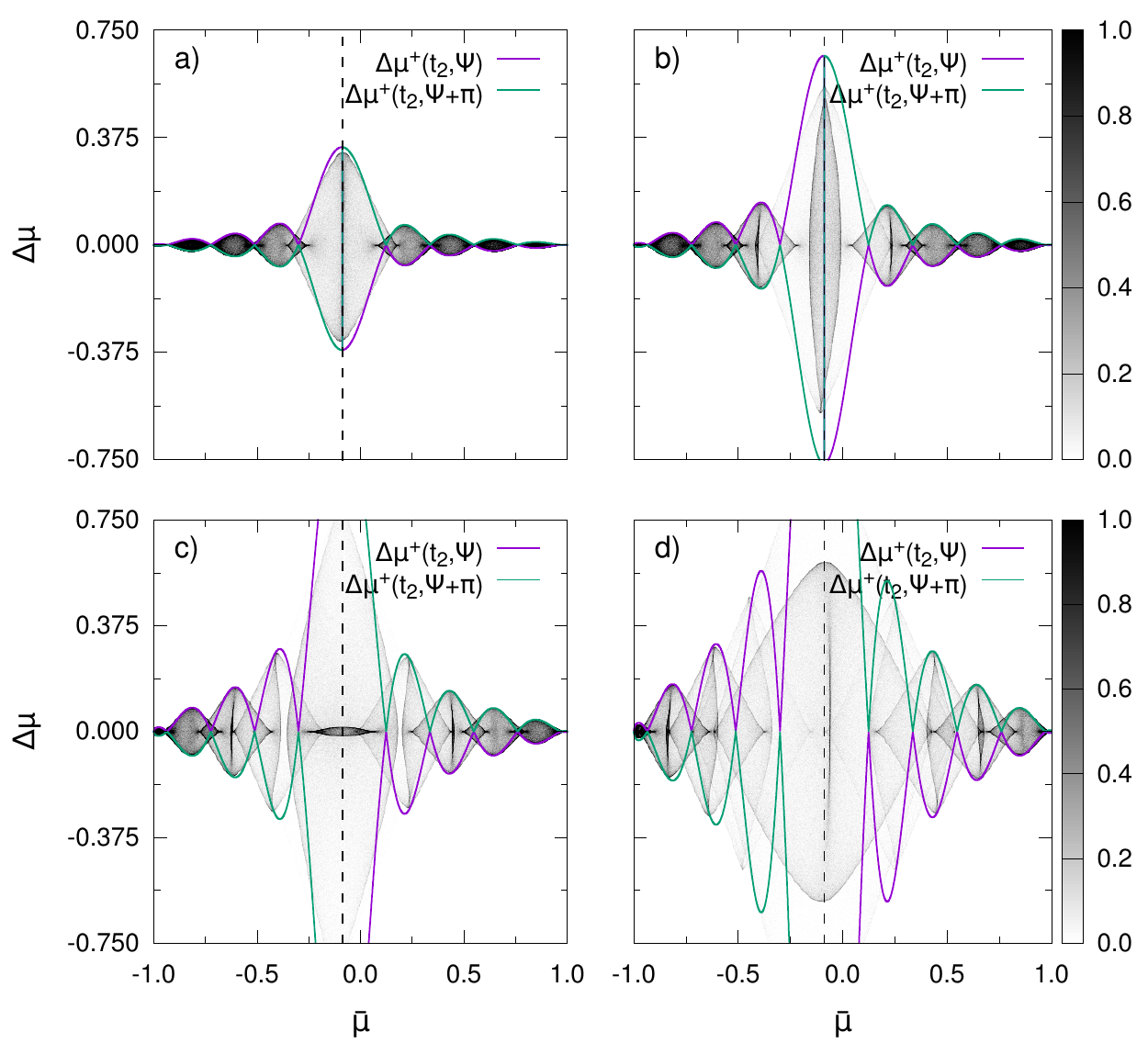}
	\caption{
		Scatter plots of electrons interacting with single plasma waves with various amplitudes, increasing from panel a) to panel d) (simulations S8 through S11).
		Simulation data is taken at time $t_2 \cdot |\Omega_\mathrm{e}| = 10.5$.
		Solid lines are analytic predictions, dashed lines mark the resonant pitch angle.
		Color coding represents test electron density in units of $10^{-4}$ of the total test particle number $N_\mathrm{e}$.
	}
	\label{fig:wave_amplitude}
\end{figure*}

In Fig. \ref{fig:wave_amplitude} four scatter plots are presented which contain data from simulations S8 through S11.
The scattering amplitude reflects the wave's amplitude as expected, as an increase of the wave's amplitude by a factor of two leads to an increase of the scattering amplitude by the same factor.
However, the asymmetry of the analytic prediction mentioned at the end of Sect. \ref{sec:plasma_frame} becomes more and more obvious with increasing amplitude.

Looking at the simulation data in Fig. \ref{fig:wave_amplitude} it can be seen that the substructure of the resonance and the ballistic peaks changes with amplitude.
This suggests that the magnetic field of the wave influences the way individual particles interact with the wave.
Especially in panels c) and d) a broadening and overlapping of ballistic peaks can also be noticed.
This effect is not recovered by the analytic approximation, which predicts that the scattering amplitude drops to zero between two neighboring peaks.
The simulation, however, shows a non-zero scattering amplitude over the whole range of pitch angles $\bar{\mu}$.

Panel d) of Fig. \ref{fig:wave_amplitude} shows a noticeable deviation of the actual resonance (as determined by the substructure of the resonance peak) and the predicted position (dashed line).
However, the dashed line is still in accordance with the position of maximum scattering amplitude.

Generally, the analytic expressions presented in Sect. \ref{sec:theory} still yield a reasonable approximation of the scattering amplitude in the case of waves with relatively large amplitudes, i.e. in the case where the assumption $\delta \! B \ll B_0$ is no longer valid.
In any case, speaking of a background magnetic field and a perturbation thereof by the magnetic field of a wave only makes sense if $\delta \! B < B_0$.
Below this limit the differences between magnetostatic QLT predictions and simulations increase for larger ratios of $\delta \! B / B_0$, but no specific amplitude can be determined at which the QLT model definitely breaks down.
This is to be expected, since QLT is derived in the limit of small perturbations of the background magnetic field and any deviation from this assumption will lead to discrepancies between the model and the actual physical system.

\subsection{Two Waves}
\label{sec:results_two_waves}

\begin{figure*}[p] 
	\centering
	\includegraphics[width=0.85\linewidth]{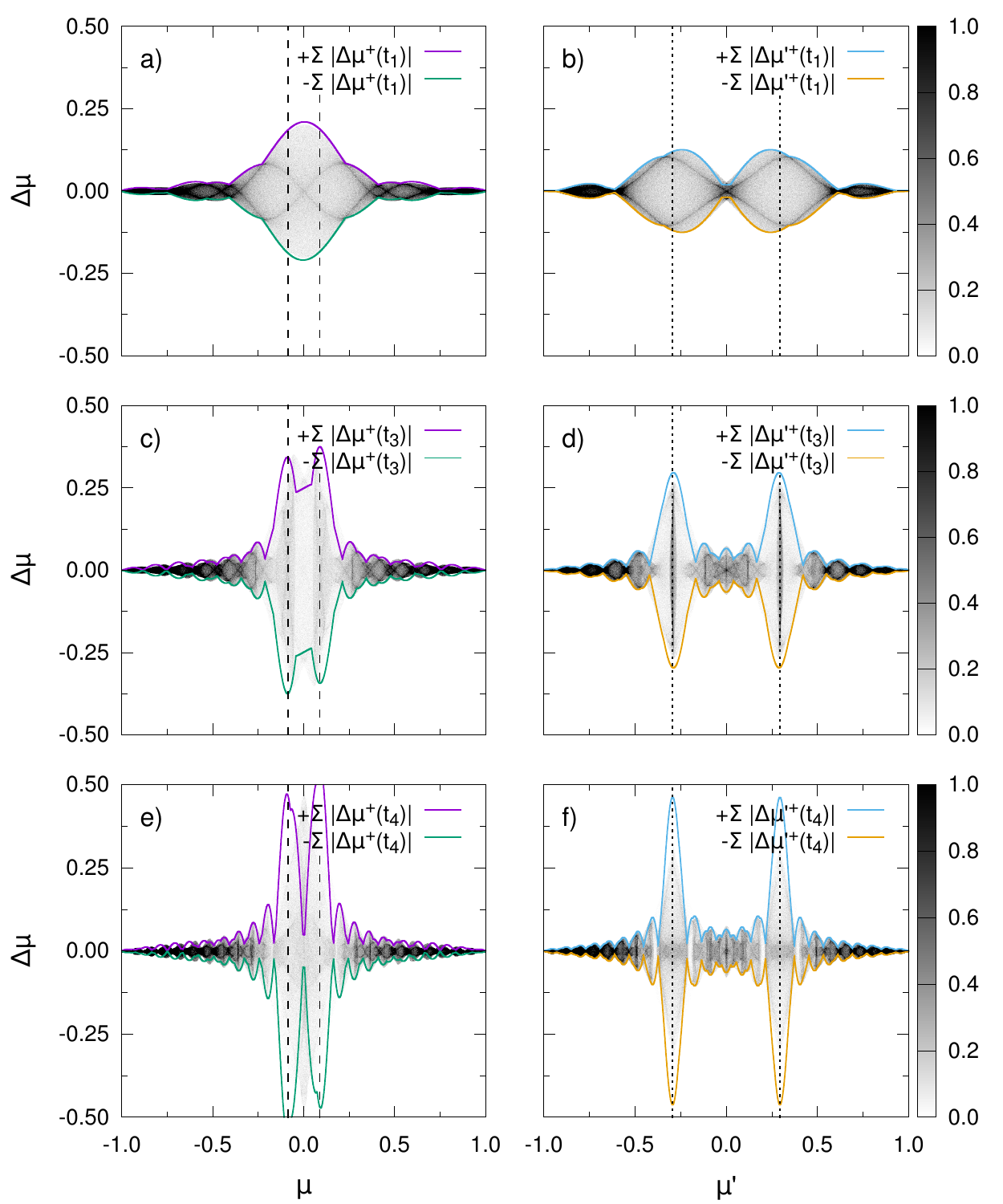}
	\caption{
		Scatter plots depicting electrons scattering off two counter-propagating waves in simulation S12 (left-hand side) and a corresponding magnetostatic simulation (right-hand side).
		Data is taken at three points in time during the simulation: $t_1 \cdot |\Omega_\mathrm{e}| = 7.0$, $t_3 \cdot |\Omega_\mathrm{e}| = 17.5$ and $t_4 \cdot |\Omega_\mathrm{e}| = 28.0$.
		Solid lines show analytic approximations for the total scattering amplitude, dashed and dotted lines mark the positions of the expected resonances in the plasma and in the wave frame, respectively.
		Simulation data is color coded as in Fig. \ref{fig:wave_amplitude}.
	}
	\label{fig:scatter_plot_two_waves}
\end{figure*}

In an attempt to study the interaction of test particles and several waves we perform a simulation (S12) which contains two waves with the same frequency and wave length, but opposite direction of propagation.
The resulting scattering amplitudes should be symmetric, as resonances lie on both sides of $\bar{\mu} = 0$ according to equation (\ref{resonance}).
In fact, this can be seen in the scatter plots shown in Fig. \ref{fig:scatter_plot_two_waves}, where the results of a full PiC simulation (S12) and a corresponding magnetostatic test particle simulation (S16) are shown.

To model the scattering amplitudes resulting from the scattering of particles and two waves we first consider the magnetostatic case.
Here we compute the $\Delta\mu'_j$ for both waves ($j = \{1,2\}$) according to equations (\ref{qlt-deltamu}) and (\ref{qlt-psi}).
Adding up the absolutes of both $\Delta\mu'_j$
\begin{equation}
	|\Delta\mu'_\mathrm{total}(\bar{\mu}')| = |\Delta\mu'_{1}(\bar{\mu}')| + |\Delta\mu'_{2}(\bar{\mu}')|
	\label{sum_scattering_amplitude}
\end{equation}
yields a curve which envelops the simulation data in the scatter plots on the right-hand side of Fig. \ref{fig:scatter_plot_two_waves}.
Thus, it seems that the total scattering amplitude is simply the sum of the scattering amplitudes for the individual waves.

In the case of the full PiC simulation a similar approach can be chosen.
Again, we can compute the scattering amplitudes $\Delta\mu_j$ for both waves using equations (\ref{qlt-deltamu}) and (\ref{qlt-psi}) and the transformations given in Sect. \ref{sec:plasma_frame}.
Note that the transformations differ for the two waves, since the waves propagate in opposite directions.
Since the resulting scattering amplitudes are not symmetric about $\Delta\mu = 0$, care has to be taken when adding up the scattering amplitudes of the different waves.
However, a sum of peace-wise defined functions $\Delta\mu(\bar{\mu})$ can be found, which maintains the asymmetries of the $\Delta\mu_j$ of each wave ($j = \{1,2\}$) and yields a representation of the total scattering amplitude.
The solid curves in the left-hand panels in Fig. \ref{fig:scatter_plot_two_waves} show the superposition of the scattering amplitudes for both waves in the plasma frame.

Contrary to the magnetostatic case, in which the analytic description matches the simulation results over the entire period of time covered by the data in Fig. \ref{fig:scatter_plot_two_waves}, the PiC simulation deviates from the prediction.
Simulation data in panel c) in Fig. \ref{fig:scatter_plot_two_waves} shows that the resonance peaks are closer to $\bar{\mu} = 0$ than theory predicts.
At even later times (panel e) the two peaks observed earlier seem to have merged into one single peak at $\bar{\mu} = 0$.

In the magnetostatic case the analytic model overestimates the amplitudes of both resonances, but peak positions agree with simulation data.
However, the resonances in Fig. \ref{fig:scatter_plot_two_waves} are much farther separated in the magnetostatic case.
An additional magnetostatic test simulation with two waves and resonances closer to $\bar{\mu}' = 0$ has been performed (not shown here).
The scatter plots obtained from this simulation are very similar to those in the left-hand panels of Fig. \ref{fig:scatter_plot_two_waves} and suggest that the superposition of the scattering amplitudes of single waves does not yield the correct total scattering amplitude even in the magnetostatic case.

These tests show that our initial assumption does not hold in general:
The total scattering amplitude resulting from the scattering of particles and several waves cannot be assumed to simply be the sum of the individual scattering amplitudes for each wave.
This becomes even more obvious in the case of another PiC simulation, S13, and its magnetostatic counterpart, S17, as we will show in the following section.

\subsection{Four Waves}
\label{sec:results_four_waves}

Simulation S13 employs the same time step length and grid spacing as the previous simulations (S8 through S12) specified in Sect. \ref{sec:simulation_setup}), but the simulation box is twice as large, with $N_\parallel = 8196$ and $N_\perp = 128$ being the number of grid cells parallel and perpendicular to $\boldsymbol{B_0}$.
The physical parameters are maintained with the exception of wave number and frequency.
Test electron energy and speed are chosen to be $E_\mathrm{kin,e} = 97.4 \, \mathrm{keV}$ and $v_\mathrm{e} = 0.54 \, c$.
Four waves are excited, as described in Table \ref{tab:phys_paras} which also lists the expected resonant pitch angles for each wave.

\begin{figure}[p] 
	\centering
	\includegraphics[width=0.75\linewidth]{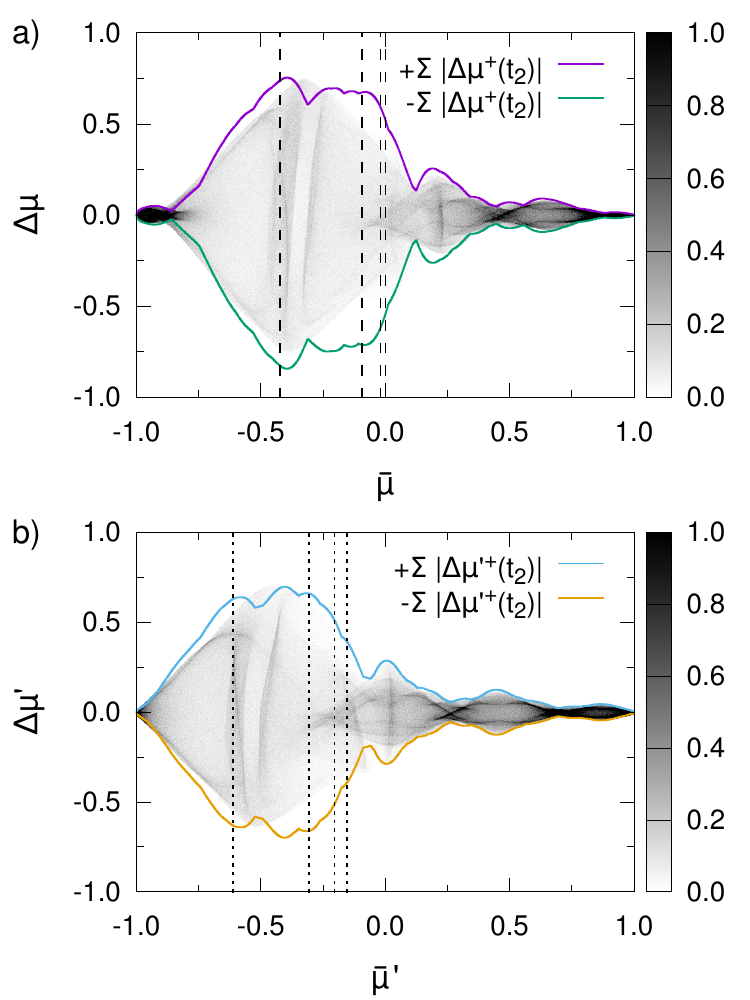} 
	\caption{
		Scatter plots from a PiC simulation (S13, top panel) and a corresponding magnetostatic test particle simulation (S17, bottom panel) with four different whistler waves.
		The plots depict pitch angle scattering at time $t_2 \cdot |\Omega_\mathrm{e}| = 10.5$.
		Solid lines represent analytic approximations for the scattering amplitudes, dashed and dotted lines mark the positions of the expected resonances for each of the four waves in the plasma and the wave frame, respectively.
		Color coding is the same as in Fig. \ref{fig:wave_amplitude}.
	}
	\label{fig:scatter_plot_four_waves}
\end{figure}

Results from both the PiC and the magnetostatic simulation are shown in Fig. \ref{fig:scatter_plot_four_waves}.
The plot illustrates that the analytic prediction of the scattering amplitude does not agree with the simulation results in either case.
The resonance peaks behave completely different than expected:
Simulation data is asymmetric about $\Delta \mu = 0$ and it does not show any of the predicted features on the edges of the resonant structure, which indicate the influence of the four different waves.
Instead, only one large blob is formed which appears roughly at the position where the resonance with the wave with smallest wave number is expected.
However, the ballistic peaks are rather well-described by theory.
Since the change of $\bar{\mu}$ is much smaller in non-resonant than in resonant scattering processes, this is to be expected as linearized theory assumes only small changes in $\bar{\mu}$ over the course of the wave-particle interaction.

\begin{figure}[p]
	\centering
	\includegraphics[width=\linewidth]{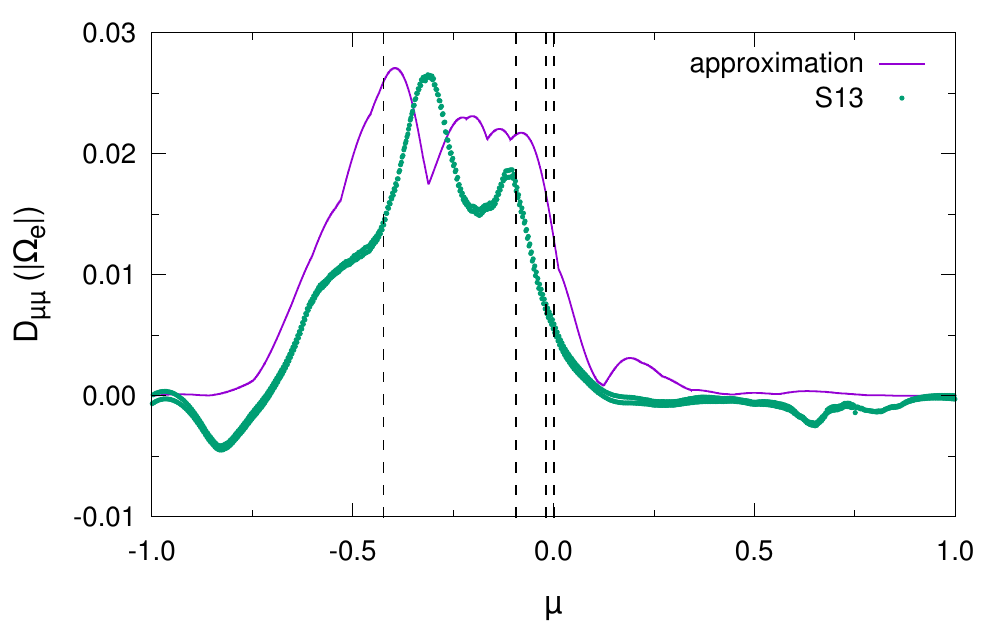}
	\caption{
		Pitch angle diffusion coefficients derived from the analytical model, Eqs. (\ref{qlt-deltamu}) and (\ref{qlt-psi}), according to the approximation in Eq. (\ref{lange_Dmumu}) and obtained from the data of simulation S13 using the method of \citet{ivascenko_2016}.
		Note that 1000 bins for $\mu$ (instead of 512) have been chosen to obtain the particle histograms.
	}
	\label{fig:Dmumu}
\end{figure}

Equations (\ref{qlt-deltamu}) and (\ref{qlt-psi}) describe the behavior of individual particles.
The results presented above demonstrate that this approach does not correctly describe wave-particle scattering in the case of several waves or large amplitudes of the waves' (combined) magnetic fields.
However, it might be possible to derive statistical properties of the system, such as the pitch angle diffusion coefficient $D_{\mu\mu}$.
We choose one of the analytical solutions for the total scattering amplitude (the one labeled ``$+\sum |\Delta\mu^+(t_2)|$'' in Fig. \ref{fig:scatter_plot_four_waves}) and derive an approximation for $D_{\mu\mu}(\mu)$ using Eq. (24) of \citet{lange_2013}:
\begin{equation}
	D_{\mu\mu} \sim \frac{(\Delta\mu)^2}{2 \, \Delta t},
	\label{lange_Dmumu}
\end{equation}
where $\Delta t = t - t_0$ is the interval between start of the interaction and the current point in time.
This approximate $D_{\mu\mu}$ is plotted in Fig. \ref{fig:Dmumu} together with a pitch angle diffusion coefficient derived from simulation data using the ``diffusion equation fitting method'' of \citet{ivascenko_2016}.
The method calculates the pitch angle diffusion coefficient from the (smoothened) test particle distributions $f(\mu,t)$ at two points in time.
Figure \ref{fig:Dmumu} shows that the simulation data produces two main features in $D_{\mu\mu}(\mu)$, which are not recovered by the approximation obtained from our analytical model.
However, the approximation recovers the correct order of magnitude, as would already be expected from looking at Fig. \ref{fig:scatter_plot_four_waves}, and even the maximum peak heights of both curves are in good agreement.

A more precise theoretical modeling of the expected diffusion coefficient was not possible, because the appropriate models \citep[e.g.][]{steinacker_1992} do not consider single waves, but spectra of waves with different frequencies and wave numbers.
Applying such a model to our test case with only four waves leads to very narrow resonance structures in $D_{\mu\mu}$, which appear to be unphysical and which do not describe the simulation data.

\section{DISCUSSION OF MICRO-PHYSICAL PITCH ANGLE DIFFUSION}
\label{sec:discussion_diffusion}

To better understand the different manifestations of the scatter plots, as seen in the previous sections, we track randomly chosen test particles throughout the simulation.
We then calculate the change of each particle's pitch angle per output interval, $\Delta\tilde{\mu} = \mu (t_n) - \mu (t_{n-1})$, where $t_{n-1}$ is the latest output time step before the current output time step $t_n$.
Plotting $\Delta\tilde{\mu}$ over the mean pitch angle $\tilde{\mu} = 0.5 \, (\mu (t_n) + \mu (t_{n-1}))$ in the interval $t_n - t_{n-1}$ yields the particles trajectory in pitch angle space.

\begin{figure*}[p] 
	\centering
	\includegraphics[width=1.\linewidth]{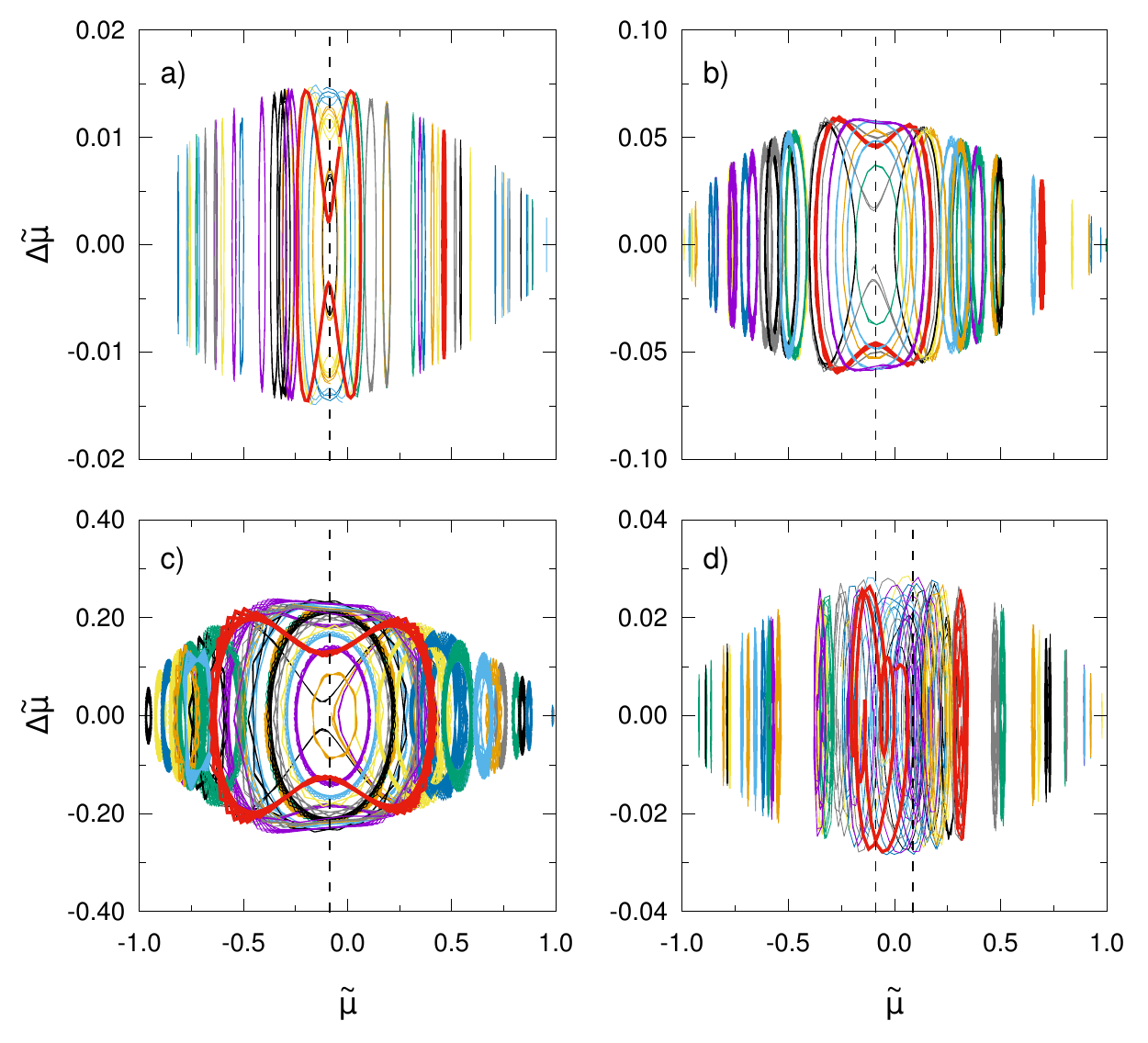}
	\caption{
		Trajectories of 50 randomly chosen test particles in $\tilde{\mu}$-$\Delta\tilde{\mu}$-space over a period of time $T \cdot |\Omega_\mathrm{e}| = 70.0$.
		The coordinates in the graph are $\tilde{\mu} = 0.5 \, (\mu(t_n) + \mu(t_{n-1}))$ and $\Delta\tilde{\mu} = \mu (t_n) - \mu (t_{n-1})$.
		Panels a), b), c) and d) show data from simulations S4, S9, S11 and S12, i.e. variations of the same simulation setup with increasing amplitude of the plasma wave (panels a through c) or an additional plasma wave (panel d).
		Each colored line represents the trajectory of one particle.
		Thick, red lines are trajectories which are discussed in the text.
		Black, dashed lines mark the position(s) of the resonance(s).
	}
	\label{fig:single_particles}
\end{figure*}

In Fig. \ref{fig:single_particles} we show such trajectories for four simulations over an interval $T$ of 80 output time steps ($T \cdot |\Omega_\mathrm{e}| = 70.0$).
The individual panels refer to four simulations:
The first is simulation S4 (panel a), which is the reference setup described in Sect. \ref{sec:results_high_energy}.
Simulations S9 and S11 (panels b and c), which share the same basic setup, contain waves with increased amplitude as discussed in Sect. \ref{sec:results_amplitude}.
Finally, simulation S12 (panel d) contains two counter-propagating waves (see Table \ref{tab:phys_paras} in Sect. \ref{sec:simulation_setup} for details on all simulations).
The respective scatter plots can be found in Figs. \ref{fig:scatter_plot_elec3}, \ref{fig:wave_amplitude} and \ref{fig:scatter_plot_two_waves}.

Scattering of electrons in simulation S4 is well described by magnetostatic QLT.
Looking at the particle trajectories in Fig. \ref{fig:single_particles} a) it can be seen that most particles' trajectories describe circles or ellipses (most of them being very narrow in terms of $\tilde{\mu}$).
The minor radii of these trajectories decrease with the distance to the resonance (e.g. the very narrow ellipsis near $\tilde{\mu} = 0.5$ marked in red).
Some trajectories that cross the resonance form a different shape which looks like the coalescence of two ellipses (one on each side of the resonance).
One of those trajectories is marked in red in Fig. \ref{fig:single_particles} a).

With increased amplitude of the plasma wave, the trajectories change only quantitatively.
Relative to S4 simulation S9 features an increase of the wave's amplitude by a factor of 4.
An increase of the amplitude $\Delta\tilde{\mu}$ by the same factor can be found when comparing Fig. \ref{fig:single_particles} b) to panel a) of the same figure.
Particles' trajectories in panel b) again form mostly circles and ellipses with a few exceptions, as described above (see trajectories marked in red).
Increasing the wave's amplitude even further (Fig. \ref{fig:single_particles} c) still yields a similar picture, although the scatter plot in Fig. \ref{fig:wave_amplitude} started to develop features which are no longer described by our analytic model.
The elliptical trajectories look distorted or polygonal, but this is only an effect of the output interval.
Otherwise, no new features can be found.
Particle trajectories in panels a), b) and c) of Fig. \ref{fig:single_particles} are rather deterministic than diffusive, since each particle seems to come back to its starting point at some time.

The most interesting case is simulation S12, containing two plasma waves.
Figure \ref{fig:single_particles}~d) again shows elliptical trajectories, but in the region around and between both resonances different forms of trajectories exist.
One of the red lines (near $\tilde{\mu} = 0.3$) has a generally elliptic shape.
However, the trajectory seems to criss-cross the ellipsis several times in an uncontrolled manner.
The other red line ($-0.25 < \tilde{\mu} < 0.1$) marks one of the cases, where the particle's trajectory is not closed at all.
The particle seems to travel along a spiral in $\tilde{\mu}$-$\Delta\tilde{\mu}$-space.
Here we assume that micro-physical diffusion sets in.
While in the previous cases most particles traveled only little along $\tilde{\mu}$, i.e. remained at their initial pitch angle, some particles can now be scattered far across the resonances and to different regimes of $\tilde{\mu}$.

We have seen that the analytic model described in Sect. \ref{sec:theory} yields a reasonable description of pitch angle scattering in the case of single plasma waves and low amplitudes.
This can be seen in the scatter plots presented in Sects. \ref{sec:simulations_1} and \ref{sec:simulations_2} which suggest that wave-particle scattering in the kinetic simulations is QLT-like.
Considering these results, the findings described above are even more interesting.
Contrary to the idea of diffusion, the individual test particles seem to be on closed trajectories if only one plasma wave is present.
The length of their trajectory as well as the pace at which the particles propagate along their trajectories depends on their initial pitch angle.

The behavior of the particles can be understood by considering the plasma wave as a series of potential wells.
\citet{sudan_1971} show that energetic electrons can be trapped in the potential wells created by whistler waves.
Assuming parallel propagation of the wave the electrons obey the parallel equation of motion in the wave frame \citep[][Eq.~(7)]{sudan_1971}:
\begin{equation}
	\frac{d v'_{\parallel \, \mathrm{e}}}{d t'} = \frac{\delta \! B'}{B_0} \, \frac{|\Omega_\mathrm{e}|}{\gamma'_\mathrm{e}} \, \sin(k'_\parallel \, z' \, t + \phi'),
	\label{equation_of_motion_trapped}
\end{equation}
where $z'$ is the spatial coordinate along the background magnetic field and $\phi'$ is the azimuth angle of the particles.
The above equation shows that potential is fluctuating in space, thus creating potential wells in periodic intervals.
In can be shown that particles whose parallel speed is $v'_{\parallel \, \mathrm{e}} = \Omega_\mathrm{e} / (\gamma'_\mathrm{e} \, k'_\parallel)$ are trapped in the potential wells.
This condition for $v'_{\parallel \, \mathrm{e}}$ corresponds exactly to the resonance condition given in Eq.~(\ref{resonance_static}), meaning that the resonant particles are trapped, whereas non-resonant electrons can escape the potential wells.
Inside their potential wells the trapped particles may oscillate back and forth along the wave.

The oscillation of a particle in the potential well corresponds to an interval in parallel speed which is given by \citep{sudan_1971}
\begin{equation}
	\Delta v'_{\parallel \, \mathrm{e}} \approx 2^{3/2} \, \sqrt{\frac{\delta \! B'}{B_0} \, \frac{\Omega_\mathrm{e}}{\gamma'_\mathrm{e}} \, \frac{v'_\mathrm{e} \, (1-\mu'^2)^{1/2}}{k'_\parallel}}.
	\label{oscillation_width}
\end{equation}
This can be translated to an interval width in pitch angle space by dividing by $v'_\mathrm{e}$.
Equation~(\ref{oscillation_width}) can be evaluated at the position of the resonance by inserting $\mu' = \mu'_\mathrm{res}$ and $v'_\mathrm{e} = v'_\mathrm{e}(\mu'_\mathrm{res})$.
Transforming the result into the plasma frame yields an interval $\Delta \tilde{\mu}_\mathrm{trap}$ which agrees well with the maximum width (along the abscissa) of those closed particle orbits in Fig.~\ref{fig:single_particles} a), b), c) which are centered around the resonant pitch angle cosine ($\Delta \tilde{\mu}_\mathrm{trap} = \{0.23, \, 0.46, \, 0.93\}$, respectively).
This supports the idea that the resonant interaction of particles and a single wave does not lead to diffusion, but rather to a trapping of resonant particles in potential wells (i.e.~closed orbits in a limited region of phase space).
Non-resonant particles are not affected by trapping effects, which is in agreement with ballistic transport.

Only in the case of two counter-propagating waves (Fig. \ref{fig:single_particles} d) or four waves with different wave numbers and frequencies (not shown here), the trajectories of single particles change in shape, which might be interpreted as pitch angle diffusion on a micro-physical level.
This suggests that several waves are required for the development of microscopic diffusion.
As pointed out in Sect. \ref{sec:note_on_single_waves} these waves need to differ not only in their phase angle, but also in frequency or wave number.
Particle transport in a turbulent medium, such as the solar wind, may therefore very well produce pitch angle diffusion on the level of individual particles.
Our simple test simulations, however, may not.

\section{CLOSING REMARKS}
\label{sec:closing_remarks}

\subsection{Summary of Results}
\label{sec:summary}

In the first part of the article (Sect. \ref{sec:theory}) we have presented an analytic model derived in the framework of QLT in the magnetostatic limit which allows to make predictions of the scattering amplitude $\Delta\mu$ as a function of a particle's mean pitch angle $\bar{\mu}$.
We have described rules of transformation which might be used to transform the model equations from the wave frame, where the magnetic fields are static, into the plasma frame, i.e. a rest frame in which the plasma wave is propagating and electric and magnetic fields change with time.
Although our approximate transformed equations yield a result which was not expected, the asymmetry of positive and negative scattering amplitudes, the overall outcome of the transformations is satisfactory.

We have then investigated electron scattering off a single whistler wave in PiC simulations and compared the numerical results to the analytic predictions of the magnetostatic QLT model.
In a first study (Sect. \ref{sec:simulations_1}) we have varied the energy of the scattering test electrons.
The qualitative shape of the scattering amplitudes as a function of the mean pitch angle remains similar over the whole energy range $1\,\mathrm{keV} \leq E_\mathrm{kin,e} \leq 1\, \mathrm{MeV}$.
However, evolution time scales and peak amplitudes change as expected.

Next, the influence of the amplitude of the magnetic field of the plasma wave has been studied (Sect. \ref{sec:simulations_2}).
Again, simulation results and analytic predictions have been compared.
Since QLT works with the assumption that the wave's amplitude $\delta \! B$ is negligible compared to the background magnetic field $B_0$, it could be expected that the analytic model fails to describe simulations with larger $\delta \! B$.
We have shown that this is the case, although the model still yields a surprisingly good approximation.
However, the magnetostatic QLT model fails to describe the interaction of particles and several waves correctly, as further simulations have shown.

In a study of the trajectories of individual particles in a simulation with only one plasma wave we have found that particles follow closed trajectories in pitch angle space, i.e. that their motion is deterministic rather than diffusive (Sect. \ref{sec:discussion_diffusion}).
This situation only changes if several plasma waves with different wave numbers or frequencies are included.

\subsection{Conclusions}
\label{sec:conclusions}

In the fully kinetic PiC simulations the propagating waves lead to changes of the absolute momenta $p$ of the individual particles in the plasma frame.
This momentum diffusion is not included in the magnetostatic QLT model, where $\dot{p} = 0$ is assumed.
However, a comparison of magnetostatic and full PiC simulations, as well as an analysis of the change of particle energy over time in a PiC simulation have shown that momentum diffusion plays only a minor role.
The assumption of constant particle energy, which was a prerequisite for the analytic expressions developed by \citet{lange_2013}, can still be applied in the electrodynamic model of PiC simulations.
This explains why the electric fields can still be neglected and why the magnetostatic model describes PiC results reasonably well after the transformation of the model equations into the plasma frame has been performed.

We found that the analytic model has difficulties in describing the effect of several waves on particle scattering.
In Sects. \ref{sec:results_two_waves} and \ref{sec:results_four_waves} we present data from simulations with two and four plasma waves.
It can be seen that the predicted scattering amplitudes only match the simulation data if the resonances for the different waves are well-separated in pitch angle space.

Looking at single particles in pitch angle space, as illustrated in Fig. \ref{fig:single_particles} in Sect. \ref{sec:discussion_diffusion}, it can be seen that in the case of single waves (or sufficiently separated resonances) each particle travels along a closed, elliptical trajectory.
In the case of single plasma waves we therefore conclude that a particle starting at some pitch angle $\mu_0$ will be scattered to different pitch angles, but will always come back to $\mu_0$ at some point in time.
This is consistent with the idea of particle trapping developed by \citet{sudan_1971}:
The magnetic field of the wave leads to effective potential wells in the rest frame of the wave.
Resonant particles are trapped and oscillate back and forth inside these potential wells, without being able to escape.
This prevents pitch angle diffusion, although the global shape of the particle distribution in phase space may change in accordance with the QLT picture of pitch angle scattering (see e.g.~Fig.~\ref{fig:scatter_plot_elec3}).

As we demonstrate in Sect. \ref{sec:discussion_diffusion}, several waves with different phase speeds or directions of propagation lead to pitch angle diffusion on a micro-physical level.
If several different waves are present particles may leave their closed trajectories and diffusion sets in.

\subsection{Context and Outlook}
\label{sec:context}

The simulations carried out and presented in this article cover the relevant energy range of electrons that interact with right-handed waves in the solar wind.
As can be seen from the model parameters from \citet{vainio_2003} and the dispersion relation for parallel propagating waves, equation (\ref{disp}), most electrons will resonate with waves in the dispersive regime.
E.g., at a distance of 2.5 solar radii electrons with energies below  $1\,\mathrm{MeV}$ will resonate with dispersive waves.
This means that for most electrons traveling from the Sun to Earth dispersive effects will play a major role and whistler waves are responsible for the resonant scattering of less energetic electrons and, therefore, cannot be neglected.
In the interstellar medium this situation is more relaxed: Here only electrons below 100 keV are affected by dispersive effects.

While the interaction of energetic particles and a spectrum of non-dispersive waves can be studied analytically (which is typically already a complicated matter), the attempt to do the same thing for dispersive waves entails a whole number of new problems.
One major difference is that wave-particle interaction can no longer be treated in the rest frame of a single wave, because different waves of the spectrum propagate at different speeds.
In the work of \citet{steinacker_1992} and \citet{vainio_2000} the QLT approach, i.e. an ensemble of low amplitude waves and the random phase approximation, has been chosen and combined with the full dispersion relation of cold plasma waves.
The resulting system of equations allows to compute the Focker-Planck coefficients $D_{\mu\mu}$, $D_{\mu p}$ and $D_{pp}$ for pitch angle and momentum diffusion in a field of dispersive waves.
\citet{vainio_2000} also gives analytic expressions for the expected resonant frequencies depending on the energy of the scattering particles.

The model for pitch angle scattering described in Sect. \ref{sec:plasma_frame} ultimately leads to a similar picture, however with a different starting point.
Other than \citet{steinacker_1992} and \citet{vainio_2000} we have started with the magnetostatic model of \citet{lee_1974} and \citet{lange_2013}, which completely neglects the effects of dispersion or electric fields.
By means of variable transformations we have made the magnetostatic model applicable to the plasma frame, i.e. a physical setting in which electromagnetic fields are not static and waves have a finite frequency and are allowed to propagate.
Plugging in the cold plasma dispersion relation we are thus able to describe the resonance of particles and any dispersive or non-dispersive wave, similar to the derivation of resonant frequencies by \citet{vainio_2000}.
However, since the original model for the pitch angle scattering amplitudes \citep{lange_2013} neglects changes in particle energy, our extended model also does not recover momentum diffusion.
The application of our model is therefore limited to regimes in which pitch angle diffusion is the fastest process, which is the case for the particle energies discussed in this article \citep[in accordance with the findings of][]{steinacker_1992}.
One major difference between our model and the work of \citet{steinacker_1992} and \citet{vainio_2000} is that they have employed a statistical approach to derive diffusion coefficients, whereas we provide a prediction for the scattering behavior of individual particles.
The expressions for the scattering amplitudes provided by \citet{lange_2013} characterize the maximum scattering efficiency which can be expected, but do not give information on the behavior of the whole ensemble of particles.

With our numerical simulations we advance from the description of individual particles to a characterization of a whole population of test particles.
The test particles are initialized with random phase angles and the scattering efficiency of each particle depends on this angle.
By means of scatter plots, including the data of a large number of individual particles, a statistical view on pitch angle scattering can be obtained.
Our data supports the existence of a maximum scattering amplitude as predicted by the model of \citet{lange_2013}, but shows that a substructure can be found within the predicted boundary.
Particle density in $\bar{\mu}$-$\Delta\mu$-space is not homogeneous, but densely or only sparsely populated regions exist.
The particle densities in the different regions cannot be derived from our analytical model, but are only seen as a result of consideration of a large ensemble of particles.

We are, therefore, of the opinion that self-consistent simulations might contribute to a better understanding of scattering processes.
Since additional effects of temperature or electric fields, which are often neglected in analytic theory, might be fully considered in numerical models, simulations may hint at shortcomings in previous theory and provide more accurate predictions.
PiC simulations of a spectrum of dispersive waves and its influence on charged particle transport, as presented by \citet{gary_2003}, are therefore a promising approach to improve our understanding of transport mechanisms in the solar wind.

A spectrum of right-handed dispersive waves can be produced self-consistently from only a few artificially amplified waves at the beginning of the simulation \citep{gary_2009,chang_2013}.
Based on this method and the results presented in the article at hand, we plan to perform further simulations of electron transport then using a full spectrum of dispersive turbulence.
Further investigations have to show, whether a two-dimensional setup is sufficient to model energetic electron transport in a turbulent medium containing dispersive waves.
Considering \citet{lange_2012_b,lange_2013}, who presented MHD simulations of Alfv\'enic turbulence and proton scattering, we expect that a three-dimensional setup is required for the correct reproduction of turbulence.
Therefore, future PiC simulations have to be designed carefully in order to keep the computational effort at a reasonable level.

\acknowledgments

The authors thank the anonymous referee for helpful comments and discussion.
The referee's knowledge of the subject and interest in our work helped to significantly improve this article.

We acknowledge the use of the \emph{ACRONYM} code and would like to thank the developers (Verein zur F\"orderung kinetischer Plasmasimulationen e.V.) for their support.

Simulations have been carried out using smaller local clusters and the resources provided by the Centre for High Performance Computing (CHPC) in Capetown, South Africa, and the SuperMUC system in Garching, Germany.
The authors gratefully acknowledge the Gauss Centre for Supercomputing e.V. (www.gauss-centre.eu) for funding this project by providing computing time on the GCS Supercomputer SuperMUC at Leibniz Supercomputing Centre (LRZ, www.lrz.de).

This work is based upon research supported by the National Research Foundation and Department of Science and Technology.
Any opinion, findings and conclusions or recommendations expressed in this material are those of the authors and therefore the NRF and DST do not accept any liability in regard thereto.

\end{document}